\documentclass[twocolumn,english,aps,superscriptaddress,pra]{revtex4-1}

\usepackage{array}
\usepackage[latin9]{inputenc}
\usepackage{amsmath}
\usepackage{amssymb}
\usepackage{graphicx}
\usepackage{babel}
\usepackage{physics}
\usepackage{mathrsfs}
\usepackage{amsfonts}
\usepackage{epstopdf}
\usepackage{xcolor}
\usepackage{mathtools}
\usepackage{multirow}
\usepackage{color}
\usepackage{natbib}

\begin{document}
\title{An Interaction Bulk-Boundary Relation\\ and its Applications Towards Symmetry Breaking and Beyond}

\author{ Saran Vijayan and Fei Zhou}
\affiliation{Department of Physics and Astronomy, University of British Columbia, 6224 Agricultural Road, Vancouver, BC, V6T 1Z1, Canada}

\begin{abstract}
In this article, we propose a simple but general scaling relation between interactions in a gapped bulk topological matter and gapless interacting surface states. 
We explicitly illustrate such a generic bulk-boundary relation ({\em BBR})
for a few specific interactions in a topological quantum matter, where we can perform dimensional reduction of a microscopic bulk theory to project out interacting surfaces. We have examined renormalization effects of the gapped bulk fermions on the interacting topological surface fermions. As simple applications, we utilize effective interacting quantum fields implied by  {\em BBR} to explore feasibility of routes to various fascinating emergent phenomena
on surfaces including emergent Majorana fermions induced by spontaneous symmetry breaking.
We obtain sufficient conditions for these interacting surface phenomena to take place. We have also found that for given bulk electron-phonon interactions and when $\Omega_D \geq m$,
the phonon-mediated interactions on surface are strongest if the bulk Debye frequency $\Omega_D$
matches $m$, the mass gap of the topological matter. 
\end{abstract}

\date{\today}
\maketitle

\section{Introduction}

A hallmark manifestation of topology in quantum matter is the existence of robust or protected boundary states in many topological quantum matter.
One very well-known one is conformal chiral Luttinger liquids along edges of incompressible fractional quantum Hall states (FQHs)\cite{Wen1, Wen2, Chang}. It is generally believed that 
in topologically ordered states such as FQHs, there is a general bulk-boundary correspondence between bulk states and conformal-field theory boundaries and a boundary
offers a unique holographic representation of complicated interacting bulks.

In relatively more recent studies of topological states\cite{Kane05, Bernevig06a, Bernevig06b, Fu07, Moore07, Qi10a, Hasan10, Qi11, Bernevig} that are usually defined in a non-interacting limit, relations between gapped bulks and boundaries become trickier because of additional axes in the parameter space
that can be further assigned to interactions. A gapped bulk with gapless surfaces imply that although interactions can be weak from a bulk perspective and do not deform the bulk matter in topological non-trivial ways,
it is possible that they substantially reshape non-interacting surfaces.
 So gapless surfaces that usually require various symmetry protections are in general much less robust to interactions because of symmetry breaking phenomena. 

For topological insulators that have been quite extensively studied over the last decade or so, if the symmetry breaking is explicit, say either due to magnetic coating or coating of a superconducting film that breaks the U(1)
symmetry, surfaces can be gapped and belong to  a different class of boundary states\cite{Fu08,Lutchyn10,Lutchyn11}. In fact, gapless surfaces can be thought to be quantum critical near zero magnetic or zero tunneling of $U(1)$ symmetry breaking fields;
and there is a surface quantum criticality that one can associate with tuning parameters that break various protecting symmetries.

If all protecting symmetries are fully respected by external fields, interactions can in principle still lead to spontaneous symmetry breaking on surfaces without deforming the gapped bulk in a non-trivial way.
The one-to-one bulk-boundary correspondence, i.e. the {\em holographic principle} suggested in topological ordered FQHs therefore does not directly apply here.  For instance, at the simplest level, one can imagine
that a weakly interacting bulk can have either a standard weakly interacting gapless surface respecting all protecting symmetries, or have a strongly interacting surface that spontaneously breaks 
one of the protecting symmetries without coated external substrates. In the latter case, topological surfaces are in a completely different phase so the same bulk can have at least two very different boundaries further depending
on interactions, especially surface interactions.

It is fascinating that in theory, surfaces can be even topologically ordered and further support fractionalized emergent new particles if all the protecting symmetries are respected and if there is no
spontaneous symmetry breaking\cite{Fidkowski,Metlitski,Wang1,Wang2,Song}. The microscopic origin of such exotic states hasn't been fully understood.
Numerical simulations of the surface model suggest that it is unlikely to realize those states with simple local interactions\cite{Neupert}. Nevertheless, if they occur in topological surfaces, they must be driven by interactions, possibly long-range ones.

One very attractive feature of weakly or non-interacting gapless boundaries, either $1D$ edges or $2D$ surfaces are that it does not have the standard Fermion-doubling as in a corresponding bulk 
$1D$ or $2D$ respectively\cite{Wu}. If ones' goal is to utilize very unique boundaries of this type as building blocks toward the more fascinating and exotic quantum matter with variable holographic features, then
perhaps one key practical step to take is to further understand how topological surfaces are interacting. Specifically, as all surfaces are boundaries of a bulk, one shall anticipate there is a general principle
that connects boundary interactions and bulk ones, hence a bulk-boundary relation ({\em BBR}) of interactions. In this article from now on, we will reserve the notion of {\em BBR} for this purpose 
only and focuses exclusively on the relations between bulk and surface interactions. 

The main object here is to illustrate a simple and practically very useful {\em BBR}  that one can apply to construct effective field theories of interacting surface fermions given the inputs of bulk interactions.
Specially, we will establish concrete relations between interaction constants in the effective field theories of surface fermions and the bulk interactions that are assumed to be known.

The article is organized as follows. In Sec. II, we will put forward a hypothesis {\em BBR} of  which relates surface interactions to bulk ones. The hypothesis is based on a simple phenomenology 
that all surface interactions shall uniquely and explicitly depend on bulk interaction constants.
When implemented in topological states, one further notices that an energy gap of bulk states, $m$, above which the bulk spectrum lies naturally defines an infrared scale (IR) of bulk dynamics.
Therefore at scale  $m$, bulk interactions that are usually introduced at a much higher ultraviolet scale associated with the bulk bandwidth $W (\gg m)$
follows an infrared renormalization flow induced by a set of {\em Renormalization Group equations} ({\em RGE}).

On the other hand,  the mass scale $m$ also defines the spatial extend of surface states into the bulk interior and can be treated as a characteristic scale of the momenta perpendicular to the surface. Therefore, $m$ appears as a natural intermediate energy scale of gapless surface states around which surface particles appear and interact. Interacting surface fermions at this scale $m$ can differ very little from bulk fermions interacting at the same scale.

\begin{figure}
\includegraphics[width=8cm]{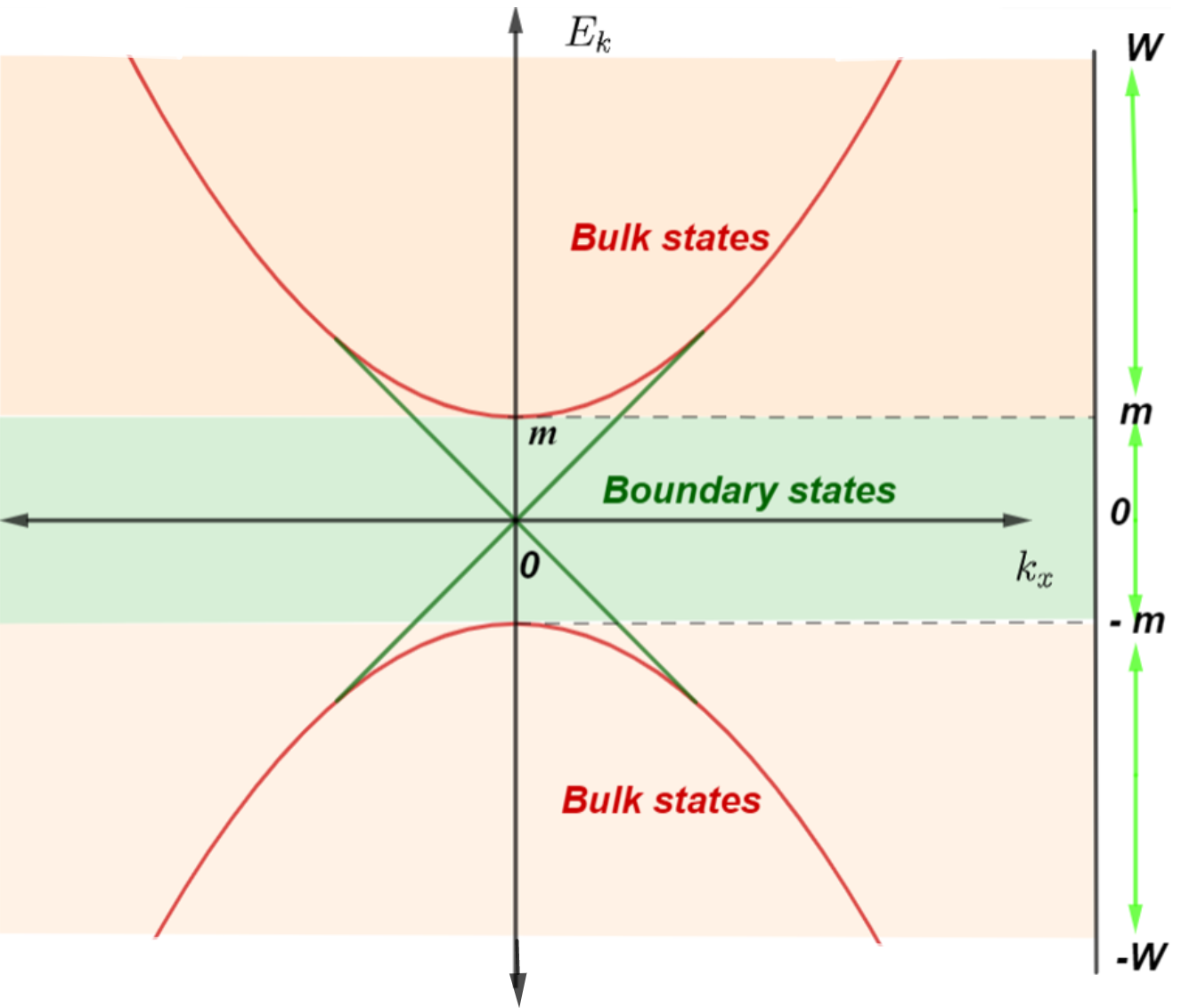}
\caption{ \footnotesize Schematic of a portion of the energy spectrum of electrons in the bulk(Brown lines) and the boundary(Green lines) in a gapped topological insulator. The mass parameter $m$ defines an energy gap of bulk states. Bulk electrons lie in the range $\Lambda_{b} \in [m, W]$, where W is the bulk UV cut-off scale. The boundary states are gapless and low energy boundary electrons of our special interests are approximately within the range of $\Lambda_{s} \in [0, m]$. The velocity of fermions has been set to be unity.}
\label{TIspectrum}
\end{figure}

We anticipate that the IR flow of {\em RGE} defined in the bulk can thus be matched with surface interactions defined at scale $\Lambda_s = m$ for low energy gapless  surfaces(see Fig.\ref{TIspectrum}).
This general matching condition
can be employed to quantify the principle of {\em BBR} and construct interaction constants in the effective field theories of gapless surface fermions. For a few concrete interactions, we derive  specific relations between bulk and boundaries.
In Sec. III, we further apply a {\em microscopic} dimension reduction technique in a few limiting cases to illustrate the general principle  and offer more quantitative features of the concrete relations. 
In Sec. IV, we discuss a specific application of our results to surface spontaneous symmetry breaking phenomena that lead to surface topological superconductivity and emergent Majorana fermions. We further apply the same idea to investigate a possibility of making supersymmetry holographic matter using interacting surfaces.

\section{General Phenomenology and Matching Conditions}

As surface states form boundaries of a bulk, it is tempting and possible to relate effective surface fields to the microscopic bulk interactions in a generic way.
Let us consider a general Hamiltonian

\begin{eqnarray}
{\mathcal H}&=&{\mathcal H}_{bulk} +{\mathcal H}_{surf}; \nonumber \\ 
{\mathcal H}_{bulk}&=&{\mathcal H}_{bulk} (\{ G_i\}, i=1,..,N; W),  \nonumber \\
{\mathcal H}_{surf} &=& {\mathcal H}_{surf} (\{ g_j\}, j=1,..,M; W_s) \nonumber \\
\end{eqnarray}
where ${\mathcal H}_{bulk(surf)}$ are the Hamiltonian acting on bulk (surface) states.
$\{ G_i\}, i=1,...,N$ and $\{ g_j\}, j=1,...,M $ are two sets of interaction constants specifying bulk and surface interactions respectively.

As stated before, $W$ is an {\em UV} scale for bulk physics which can be naturally associated with the {\it bandwidth}. $W_s$ is an {\em UV}  scale of surface states which further depend on surface band microscopic structures as well as types of interactions of interests. And typically, both $W$ and $W_s$  are much bigger than the mass gap $m$.

The mass gap plays a paramount role in our discussions of {\em BBR}. For topological states we will focus on, $m$ can be thought as an IR scale of bulk physics above which all bulk states lie.
On the other hand, the mass gap $m \in [0, W_s] $ simultaneously is also a characteristic intermediate energy scale of gapless surface dynamics because $\frac{1}{m}$ is set by the spatial spread of surface states into the bulk interior. 
So indeed at this particular scale of $m$, two dimensional surface states appear to be no different from three dimensional bulk ones at the same scale.  
The matching between the {\em IR} physics of the interacting bulk and the physics of interacting gapless surface states at this special intermediate scale of $m$ is a key observation that leads to the principle of interaction {\em BBR}.

\begin{figure}
\includegraphics[width=6cm]{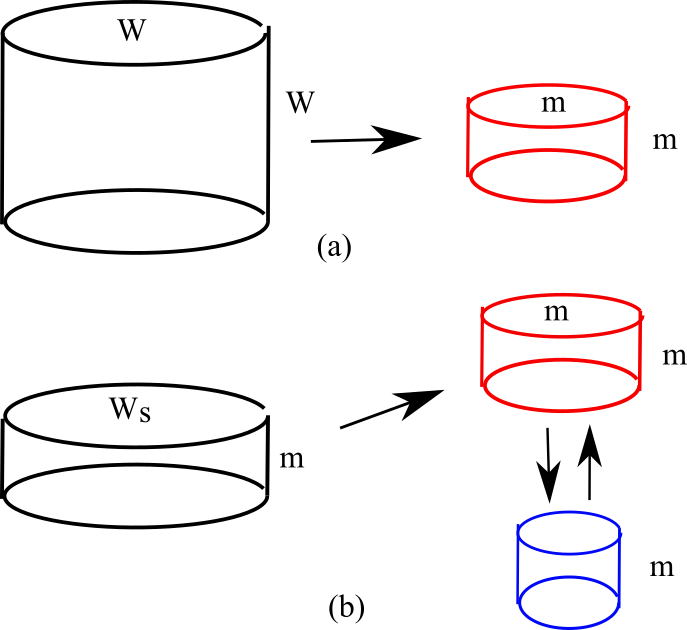}
\caption{ \footnotesize A cartoon of 
a) bulk scale transformation from an $UV$ scale $\Lambda_b=W$ to the bulk infrared scale $\Lambda_b =m$ (red online) and b) surface scale transformation 
from an {\em UV} scale $\Lambda_s=W_s$ to $\Lambda_s=m$ (red online) to an infrared scale $\Lambda_s \ll m$ (blue online) of surface fermions in a 3D momentum space.  In b), the scale transformation is performed only in the two dimensional horizontal plane while the vertical momentum remains fixed at a scale
similar to $m$.
a) and b) suggest a bulk-boundary matching condition.
The effective theories obtained in a) and b) at scale of $m$  (red online) appear to be equivalent and so can be matched to obtain a {\em BBR} relation between bulk and surface interactions. (See the main text for details of {\em BBR}). }
\label{BBR}
\end{figure}

There is an alternative and perhaps simpler view of {\em BBR} . For certain discussions, it is even possible to treat $m$ simply as an effective {\em UV} scale of interacting gapless surface fermions or $W_s \sim m$. And we can let the scale transformation generated by {\em RGE}s of surface fermions run from a surface IR  scale of  nearly zero energy, with some pre-assumed interactions, toward some ultraviolet energies rather than the other way around. In that case, we can conveniently match the {\em IR} oriented flow of the bulk interactions with the {\em UV} oriented flow of the surface interactions at scale $m$. The match leads to a surface theory 
formulated at the scale $m$ with bare interaction constants determined by the matching conditions in  {\em BBR} (see Fig.\ref{TIspectrum}, Fig.\ref{BBR}).

In more generic situations where $W_s \geq m$, one can still apply {\em BBR} to construct an effective field theory at an intermediate scale $m$.
Once the matching condition has been employed at scale $m$, following the {\em UV} flow generated by {\em RGE}s (see more discussions in section IV),
one can also easily further extend the theory to the scale $W_s$, that is usually higher than $m$ to construct the full surface field theory up to the {\em UV} scale $W_s$.
Of course if one wishes, one can also reverse the flow to scales smaller than $m$ back to infrared when needed and suitable. In this case, with the effective surface field theory already fixed at the effective {\em UV} scale $W_s$ or an intermediate scale $m$, one can find out the pre-assumed near zero energy theories.

Formally speaking,
following the general scheme of renormalization and scaling, both bulk and surface fields can be equivalently redefined at an arbitrary running {\em UV} energy scale $\Lambda$. In the rest of discussions, for convenience and simplicity, 
for bulk interactions, $\Lambda$ is chosen to be larger than $m$, while for surfaces,  we restrict our discussions to energy scales lower than $m$ (although all these conditions can be relaxed in general).
The effective theories defined at $\Lambda_{b,s} $, where subscripts $b,s$ are for bulk and surface respectively, can be expressed as

\begin{eqnarray}
{\mathcal H}_{bulk} &= & \Lambda^z_b {\mathcal F}_{bulk} (\{ \tilde{G}_i (\Lambda_b), i=1,...,N \}, Z_{f}(\Lambda_{b}); \frac{m}{\Lambda_b} <1), \nonumber \\
{\mathcal H}_{surf} &=&\Lambda^z_s {\mathcal F}_{surf} (\{\tilde{g}_j (\Lambda_s), j=1,..,M \}, Z_{f}(\Lambda_{s}); \frac{m}{\Lambda_s} >1 ).\nonumber\\
\end{eqnarray}
Here we have defined $z$ as the dynamical critical exponent of our system which will be set to be $z=1$.
${\mathcal F}_{bulk,surf}$ are dimensionless and
dimensionless coupling constants $\tilde{G}_i (\Lambda_b), i=1,..,N$, and $\tilde{g}_j (\Lambda_s), j=1,...,M$ are further introduced to facilitate discussions below.
Their scale dependence on $\Lambda_{b,s}$ can be obtained via the standard renormalization group analysis and details will be presented later. $Z_{f}$ stands for the fermion field renormalization which does not play significant role in our phenomenological discussions and we will mute from now on.

An important phenomenology here is that bulk interactions defined at  $\Lambda_b=m$ in $\mathcal{H}_{bulk}$ shall describe, effectively and precisely, interacting boundary dynamics of $\mathcal{H}_{surf}$ defined at $\Lambda_s=m$.
One clear evidence for this is that surface states can merge into bulk spectrum at scale $m$ and there is no longer a clear separation between bulk and surface states at that energy scale.
Therefore, at least if in the case when $N=M$ and if all bulk interactions adiabatically evolve into surface counterparts as $\Lambda_b$ is lowered toward $m$,
one can anticipate surface interactions at that {\em UV} scale of $\lambda_{s} = m$ shall be {\em equivalent} to
bulk interactions. In this particular case, we have a rather simple relation at the IR scale $\Lambda_{b} = m$,

\begin{eqnarray}
\tilde{g}_i(\Lambda_s=m) =\tilde{G}_i(\Lambda_b=m), i,j=1,...,N=M.
\label{Match}
\end{eqnarray}

In more general cases, Eq.\ref{Match} shall be replaced by a more general relation,

\begin{eqnarray}
\tilde{g}_i(m) = \Pi_i(\{ \tilde{G}_j(m)\}, j=1,...,M), i=1,...,N
\label{Match1}
\end{eqnarray}
where $\Pi_i, i=1,..N$ are dimensionless functions that can be generated by general matching conditions. 
Below, we will further illustrate explicitly these functions in a few cases.

\begin{figure}
\includegraphics[width=8cm]{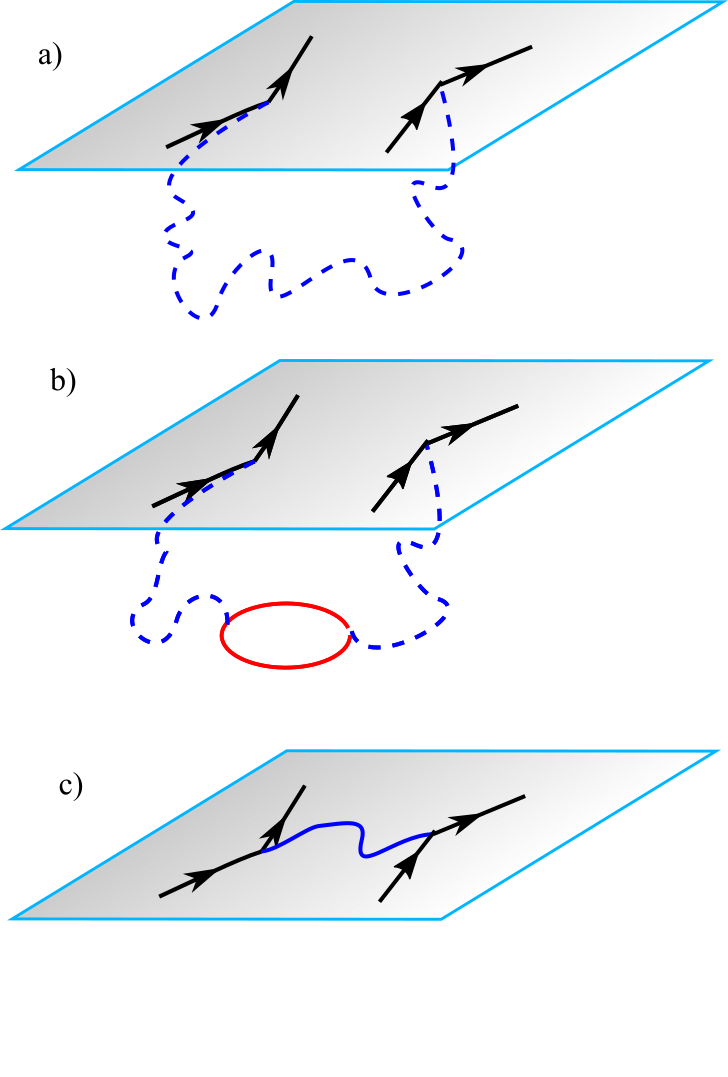}
\centering
\caption{ \footnotesize The schematic of bulk-boundary relation of interactions. 
(a) and (b) illustrate surface fermions interacting with higher energy bulk phonons or photons without or with fermion polarization effects respectively. A fermion loop
in (b) indicates general bulk renormalization effects that lead to suppression of coulomb interactions at lower energy scales. (c) illustrate effective surface interactions in the infrared limit. The bulk-boundary relation principle matches a)-b) and c) at a mass gap $m$ that effectively defines a surface band width. }
\label{bulkboundary}
\end{figure}

\begin{figure}
\includegraphics[width=8cm]{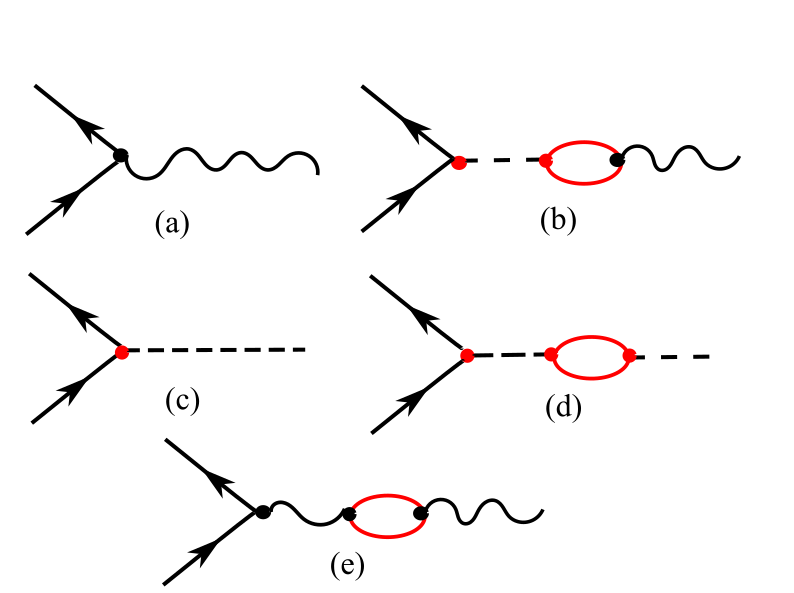}
\centering
\caption{ \footnotesize Diagrams that contribute to the renormalization group equations in Eq.\ref{RGEcph} . Solid lines are for fermions; wave lines are for phonons and dashed ones for Coulomb interactions.
a) and b) are for phonon-electron interactions and c) and d) are for Coulomb interactions; e) only contributes to the momentum dependence of fermion-phonon interactions $G_{fp}$, and does not directly contribute to  Eq.\ref{RGEcph}. (The vertex correction and fermion self-energy contributions are not shown here as they lead to less relevant terms in the infrared limit and also do not contribute to our analyses.
)}
\label{DiagramsRGE}
\end{figure}

\subsection{Surface Coulomb Interactions}

Now let us take 3D massive Dirac fermions with coulomb interactions as an effective theory for 3D bulk topological states as the simplest example.
Surfaces are then naturally described by interacting two-component helical fermions.

 \begin{eqnarray}
{\mathcal H}_{bulk} &= & \int d^3{\bf r} \psi_b^\dagger \left[v_{F} \sigma_z\otimes {\bf s}\cdot {\bf p} - m \sigma_{x} \right]\psi_b +  \nonumber \\
&+& e_{b}^2
\int d^{3}{\bf r} d^3{\bf r'} \psi_b^\dagger ({\bf r}) \psi_b({\bf r})  \frac{1}{|{\bf r} -{\bf r}'|} \psi_b^\dagger({\bf r}') \psi_b({\bf r}'), \nonumber \\
{\mathcal H}_{surf} &=&  v_{F} \int d^2{\bf r} \psi_s^\dagger {\bf s}\cdot {\bf p} \psi_s \nonumber \\
&+& e_s^2
\int d^{2}{\bf r} d^2{\bf r'} \psi_s^\dagger ({\bf r}) \psi_s({\bf r})  \frac{1}{|{\bf r} -{\bf r}'|} \psi_s^\dagger({\bf r}') \psi_s({\bf r}')
\end{eqnarray}
where ${\bf p}$ is a momentum gradient operator defined either in a bulk or on the top surface (and we only show one surface here). 
$\psi^\dagger_s$ is a creation operator of  two-component helical surface fermions that can be projected out of four component bulk operators, i.e. $\psi^\dagger_s ={\mathcal P}_+ \psi^\dagger_b$
and ${\mathcal P}_+ =(1+\tau_z)$.  Finally, $e_{s,b}$ are surface and bulk charges respectively; they are identical only at a classical level and can be drastically different once quantum renormalization
effects due to polarizations are properly taken into account.

Dirac fermions with long range Coulomb interactions had been well understood in previous studies of graphene\cite{gonzalez99,schmalian07,sdsharma,lucas}. Very similar calculations lead to simple renormalization group equations
for ${\mathcal H}_{b,s}$ that we list below.  We express them in terms of dimensionless coupling constants for the bulk (b) and surface (s) 

\begin{eqnarray}
\alpha_\eta=\frac{e_{\eta}^2}{\hbar v_{F\eta}}, \eta=b, s.\label{alphaRG}
\end{eqnarray}

In terms of these dimensionless constants, we have

\begin{eqnarray}
\frac{\partial \alpha_{\eta} }{\partial t_\eta }&=&c_\eta \alpha_\eta^2, \eta=b,s \nonumber \\
\alpha_b (t_b=0)&=&\alpha_{b0}, \alpha_s (t_s=0)=\alpha_{s0}
\label{RGEc}
\end{eqnarray}
where $t_\eta=\ln \Lambda_\eta /\Lambda_{\eta 0}$ is a standard dimensionless running scale defined in terms of $\Lambda_\eta$, $\eta=b,s$. 
$c_\eta$, $\eta=b,s$ are constants that do not carry much significance (Using Wilson-Fisher RG analysis\cite{Peshkin}, we get $c_{s} = \frac{1}{4\pi^{2}}$ and $c_{b} = \frac{5}{12\pi^{2}}$).
Note that for $\alpha_b$, $\Lambda_b$ runs from $m$ up to the ultraviolet scale $\Lambda_{b0}= W$ while for $\alpha_s$, 
$\Lambda_s$ runs below $\Lambda_{s0}=m$ because surface states are present below mass gaps.
The second line in Eq.\ref{RGEc} defines the flow initial condition for bulk and surface interactions at their corresponding ultraviolet scales of $\Lambda_{b0}=W$, $\Lambda_{s0}=m$.

In this case, there is single interaction constant. We expect bulk Coulomb interactions adiabatically evolves into surface ones at matching scale $\Lambda=m$.
Therefore,

\begin{eqnarray}
\alpha_s (\Lambda_s=m)& =&\alpha_b(\Lambda_b=m) 
\mbox{	or	} \nonumber \\
\alpha_{s0} (\alpha_{b0}, \frac{m}{W})&=&\frac{\alpha_{b0}}{1+c_b \alpha_{b0} \ln \frac{W}{m}} 
\label{RGEc1}
\end{eqnarray}
where we have solved Eq.\ref{RGEc} for bulk interactions to derive the second line.

The results in Eq.\ref{RGEc1} appear to be quite intuitive. It states that interacting surface shall be naturally characterized by the low energy effective theory of bulk interactions. 
In this particular case, surface interaction constant is simply equal to, the running coupling constant of the bulk $\alpha_{b} (\Lambda_b)$ at matching scale $m$.
As a consequence, surface interactions (dimensionless one) have to be a simple function of two dimensionless parameters: 1) bulk interaction constant $\alpha_{b0}$ and 2) surface 
"bandwidth" or band gap, $m$, measured in terms of bulk bandwidth $W$. As we will see later, this feature is a robust one and appears in all the cases we are examining. 
It is a simple but remarkable manifestation of the principle of interaction {\em BBR}.

\subsection{Charge and velocity renormalization in bulk and on surface}

Although the renormalization equations for bulk and surface coupling constants $\alpha_{b,s}$ are qualitatively similar, mechanisms leading to them are very different. 
Namely, in the bulk, both charge and velocity renormalization contribute to the renormalization of coupling constant $\alpha_b$ while on surface, charge renormalization is absent and
the renormalization of coupling constant $\alpha_s$ is solely due to the velocity renormalization. The absence of surface charge renormalization effects is generic of $2D$ Dirac fermions with Coulomb interactions
and they were studied and pointed out in previous studies of graphene physics\cite{gonzalez99, schmalian07, sdsharma, lucas}.

In the bulk, we find the velocity and charge renormalization to the one-loop approximation are given by

\begin{eqnarray}
\frac{\partial \tilde{e}_b^2 }{\partial t} = c_{be}\alpha_{b} \tilde{e}^2_b,
\frac{\partial \tilde{v}_F}{\partial t} = -c_{bv} \alpha_{b} \tilde{v}_F; c_{be}, c_{bv} >0. 
\label{ev} 
\end{eqnarray}
Eq.\ref{ev} leads to Eq.\ref{RGEc}  for $\alpha_b$ with $c_b=c_{be} +c_{bv}$. Here $c_{be} = \frac{1}{6\pi^{2}}$ and $c_{bv} = \frac{1}{4\pi^{2}}$

 One can verify the following solution for the charge and velocity,
 
 \begin{eqnarray}
 &&
 \frac{\tilde{e}^2_b}{\tilde{e}^2_{b0}}=(\frac{\alpha_{b}(\Lambda) }{\alpha_{b0}})^{A},
  \frac{\tilde{v}_F}{\tilde{v}_{F0}}=(\frac{\alpha_{b}(\Lambda) }{\alpha_{b0}})^{- B}. \nonumber \\
 && \text{where}\,\,\,A=\frac{c_{be}}{c_{be} +c_{bv}}, B=\frac{c_{bv}}{c_{be} +c_{bv}}; A, B>0\nonumber\\ && \text{and} \,\,\,\, \alpha_{b} (\Lambda_{b}) =\frac{\alpha_{b0}}{1+c_b \alpha_{b0} \ln \frac{W}{\Lambda_{b}}}.
 \end{eqnarray}
Note that $A+B=1$. Taking into account the values of $c_{be}$ and $c_{bv}$, we find that $A = \frac{2}{5}$ and $B = \frac{3}{5}$

 On surfaces, $c_{se}=A=0$ hence there is no charge renormalization. On the other hand, $c_{sv}=c_s$ and $B=1$.
The velocity increases logarithmically as low energies are approached. Similar aspects were discussed in two dimensional graphene.

\subsection{Phonon-Mediated Attractive Surface Interactions}

Next example we will focus on is possible phonon-mediated attractive surface interactions. For the sake of simplicity, in this part of discussion, we are going to assume this is the only way fermions interact and neglect other interactions such as Coulomb ones. In next subsection, we will restore other interactions.
We will show that again surface interactions can be simply related to two important aspects of topological matter
(and these are the {\em only} two factors relevant to surface interactions):
1) bulk phonon-interaction constants $\tilde{G}_{fp0}$(dimensionless) defined at a scale of phonon bandwidth or Debye frequency $\Omega_D$;
2) surface state bandwidth $m$ measured in terms of phonon bandwidth, $\frac{m}{\Omega_D}$ which we assume, without losing generality, to be less than one. 

 \begin{eqnarray}
{\mathcal H}_{bulk} &= & {\mathcal H}_{f} +{\mathcal H}_{ph} +{\mathcal H}_{I} \nonumber\\
{\mathcal H}_f &=&  \int d^3{\bf r} \psi_b^\dagger \left[ v_{F}\sigma_z\otimes {\bf s}\cdot {\bf p} - m \sigma_{x} \right] \psi_b, \nonumber \\
{\mathcal H}_{ph} &=&\int d^3{\bf r} [ {\bf \pi}({\bf r})^2 +[\nabla {\bf \phi} ({\bf r})]^2  ] \nonumber \\
{\mathcal H}_{I} &=& G_{fp}\int d^{3}{\bf r} \vec{\nabla} \cdot \vec{\phi} ({\bf r}) \psi_b^\dagger ({\bf r}) \psi_b({\bf r}) , \nonumber \\
{\mathcal H}_{surf} &=&  v_{F} \int d^2{\bf r} \psi_s^\dagger {\bf s}\cdot {\bf p} \psi_s 
\\ &+& g_1
\int d^{2}{\bf r} \psi_s^\dagger ({\bf r}) \psi_s({\bf r}) \psi_s^\dagger({\bf r}) \psi_s({\bf r}). \nonumber \\
\end{eqnarray}
$\psi_{b}$ and  $\psi_{s}$ are the bulk and the surface Dirac fields respectively. Here phonon fields are expressed in terms of ${\bf \phi}$ and ${\bf \pi}$ which form a pair of canonical dynamical fields, i.e.

\begin{eqnarray}
[{\bf \phi}_p({\bf r}), {\bf \pi}_q ({\bf r}')] =i\hbar \delta_{p,q} \delta({\bf r}-{\bf r}'). 
\end{eqnarray}

The bulk theory when casted at a running {\em UV} scale ${\Lambda_b}$  can be characterized by a dimensionless renormalization charge $\tilde{G}_{fp}(\Lambda_b) =G_{fp} \Lambda_b^{(D_b-1)/2}$ with $D_b=3$ for bulk;
surface interactions on the other hand can be characterized by $\tilde{g}_1(\Lambda_s)=g_1 \Lambda_s^{d_s-1}$ with $d_s=2$ for surfaces.  Under scale transformation, they transform according to the following equations;

\begin{eqnarray}
\frac{\partial \tilde{G}_{fp} }{\partial t_b} &=& \frac{D_b-1}{2}  \tilde{G}_{fp} \nonumber\\
\frac{\partial \tilde{g}_1}{\partial t_s} &=& \left(d_{s} - 1\right) \tilde{g}_1 +\tilde{g}_1^2 \nonumber\\
\tilde{G}_{fp} (t_b=0)&=& \tilde{G}_{fp0}, \tilde{g}_1 (t_s=0)=\tilde{g}_{10} 
\label{RGEph}
\end{eqnarray}
where again $t_b=\ln ( \Lambda_b/\Omega_D)$ and $t_s=\ln (\Lambda_s/m)$;
$\Lambda_b \in [m, \Omega]$ and $\Lambda_s \in [0,m]$.
It is worth remarking that in this limit, the one-loop contribution to the {\em RGE} of $\tilde{G}_{fp}$ in Eq.\ref{RGEph}, which is of order of $\tilde{G}_{ph}^3$, vanishes identically because the electron-phonon interaction vertex is further proportional to the momentum transfer.

As in conventional metals,
electron-phonon interactions mediate an attractive interaction between bulk Dirac fermions. We can characterize such interactions by a conventional four-fermion operator in the bulk,

\begin{eqnarray}
{\mathcal H}^{eff}_{I} &=& g_b
\int d^{3}{\bf r} \psi_b^\dagger ({\bf r}) \psi_b({\bf r}) \psi_b^\dagger({\bf r}) \psi_b({\bf r}), \nonumber \\
 \tilde{g}_b(\Lambda_b) & =&-\tilde{G}^2_{pf} (\Lambda_b). 
 \label{EFF}
\end{eqnarray}
where  $\tilde{g}_b(\Lambda_b) =g_b\Lambda_b^{D_b-1}$.

At $\Lambda_b=\Lambda_s=m$,
we then apply the following matching condition for $\tilde{g}_1(\Lambda_s=m)$, the phonon mediated interaction on surface; 
\begin{eqnarray}
\tilde{g}_{1} (\Lambda_s=m) =\tilde{g}_b (\Lambda_b=m) = - \tilde{G}^2_{fp} (\Lambda_b=m). 
\end{eqnarray}
This matching condition leads to the following relation between the surface interactions and the bulk ones,

\begin{eqnarray}
\tilde{g}_{10}(\tilde{G}_{fp0}, \frac{m}{\Omega_D}) &=& - {\tilde{G}^2_{fp0}}
(\frac{m}{\Omega_D})^{{D_b-1}}.
\label{match2}
\end{eqnarray}
One can obtain Eq.\ref{match2} by solving Eq.\ref{RGEph} with its ultraviolet boundary condition defined at $t_b=0$ or $\Lambda_b=\Omega_D$.
The structure is anticipated as surface attractive interactions are mediated by electron-phonon interactions and so $\tilde{g}_{10}$ is uniquely set by bulk characterization of
phonon fields, $\tilde{G}_{fp0}$ and relative surface bandwidth, $m/\Omega_D$.

\subsection{A more general case}

Now we turn to a semi-realistic situation where both long range Coulomb interactions and phonon interactions are present.

 \begin{eqnarray}
{\mathcal H}_{bulk} &= & {\mathcal H}_{f} +{\mathcal H}_{ph} +{\mathcal H}_{I} \nonumber\\
{\mathcal H}_f &=&  \int d^3{\bf r} \psi_b^\dagger \left[ v_{F}\sigma_z\otimes {\bf s}\cdot {\bf p} - m \sigma_{x} \right] \psi_b, \nonumber \\
{\mathcal H}_{ph} &=&\int d^3{\bf r} [ {\bf \pi}({\bf r})^2 +[\nabla {\bf \phi} ({\bf r})]^2  ] \nonumber \\
{\mathcal H}_{I} &=& G_{fp}\int d^{3}{\bf r} \vec{\nabla} \cdot \vec{\phi} ({\bf r}) \psi_b^\dagger ({\bf r}) \psi_b({\bf r}) , \nonumber \\
&+& e_{b}^2
\int d^{3}{\bf r} d^3{\bf r'} \psi_b^\dagger ({\bf r}) \psi_b({\bf r})  \frac{1}{|{\bf r} -{\bf r}'|} \psi_b^\dagger({\bf r}') \psi_b({\bf r}') \nonumber \\
{\mathcal H}_{surf} &=&  v_{F} \int d^2{\bf r} \psi_s^\dagger {\bf s}\cdot {\bf p} \psi_s 
+ g_1
\int d^{2}{\bf r} \psi_s^\dagger ({\bf r}) \psi_s({\bf r}) \psi_s^\dagger({\bf r}) \psi_s({\bf r}) \nonumber \\
&+& e_s^2
\int d^{2}{\bf r} d^2{\bf r'} \psi_s^\dagger ({\bf r}) \psi_s({\bf r})  \frac{1}{|{\bf r} -{\bf r}'|} \psi_s^\dagger({\bf r}') \psi_s({\bf r}'). \label{Hfull}
\end{eqnarray}

To determine the effective surface field theories in this case, we again match the renormalization flow of bulk interactions in the IR region near $\Lambda_b=m$ with 
the renormalization flow of surface fermions in its UV region $\Lambda_s=m$.  As discussed before, the phonon mediated interactions can be conveniently represented by an effective attractive interaction, $\tilde{g}_b$, defined Eq.\ref{EFF}.
Therefore,  {\em BBR} suggests a matching between surface interactions and bulk ones at $m$, i.e. $\tilde{g}_{1} (\Lambda_s=m) =\tilde{g}_b(\Lambda_b=m)$.  

More specifically,
\begin{eqnarray}
\alpha_{s0}=\alpha_s (\Lambda_s=m)& =&\alpha_b(\Lambda_b=m), 
\nonumber \\
\tilde{g}_{10}=\tilde{g}_{1} (\Lambda_s=m) &=& - \tilde{G}^2_{fp} (\Lambda_b=m).
\label{match3}
\end{eqnarray}

This can be easily implemented by working out similar renormalization group equations for bulk interactions $\tilde{G}_{fp}$ and $\tilde{e}^2_b$ in this case for $D_b=3$ or a $3D$ bulk.

\begin{eqnarray}
\frac{\partial \tilde{G}_{fp} }{\partial t_b} &=& \frac{D_b-1}{2}  \tilde{G}_{fp} 
+c_b \alpha_b \tilde{G}_{fp} ,
\nonumber\\
\frac{\partial \alpha_{b} }{\partial t_b }&=&c_b \alpha^2_b \nonumber \\
\tilde{G}_{fp} (t_b=\ln \frac{\Omega_D}{W} )&=& \tilde{G}_{fp0},\alpha_b (t_b=0)=\alpha_{b0}.
\label{RGEcph}
\end{eqnarray}

Note that $t_b$ is defined as $t_b=\ln \Lambda_b/W$ and so the boundary conditions for $G_{fp}$ is set at $t_b=\ln \Omega_D/W \neq 0$.
And we have focused on the limit where $\Omega_D \ll W$ although one shall find it straightforward to generalize to other limits.

The solutions to Eq.\ref{RGEcph}  can be obtained and matching conditions in Eq.\ref{match3} lead to

\begin{eqnarray}
\alpha_{s0} (\alpha_{b0}, \frac{m}{W})&=&\frac{\alpha_{b0}}{1+c_b \alpha_{b0} \ln \frac{W}{m}} \nonumber \\
\tilde{g}_{10}(\tilde{G}_{fp0}, \frac{m}{\Omega_D}) &=&\frac{-\tilde{G}^2_{fp0}}{\left[1+c_b \alpha_{b} (\Omega_D) \ln \frac{\Omega_D}{m}\right]^{2} }
(\frac{m}{\Omega_D})^{D_b-1}. \nonumber \\
\label{match4}
\end{eqnarray}
When presenting $\tilde{g}_{10}$ in Eq.\ref{match4}, we have only kept the lowest order contributions in terns of $\tilde{G}_{fp0}$. 

It is worth emphasizing that two different coupling constants entering Eq.\ref{match4}: 1) $\alpha_{b0}$, the bare {\em coupling constant} defined at $\Lambda_b=W$, the bulk {\em UV} scale, which results in the effective surface charges
$\alpha_{s0}$ and  2) $\alpha_b(\Omega_D)$ that is defined as

\begin{eqnarray}
\alpha_b(\Omega_D) =\frac{\alpha_{b0}}{1+ c_b \alpha_{b0} \ln \frac{W}{\Omega_D} },   
\end{eqnarray}
the renormalized {\em coupling} defined at Debye frequency $\Omega_D$.
And $\alpha_b(\Omega_D)$  is much less than its bare value $\alpha_{b0}$ if $ \ln (W/\Omega_D) \gg 1$  and $\alpha_{b0}$ is of order unity as in solid states.
The later {\em coupling} contributes to attractive interaction strength $\tilde{g}_{10}$  that is further renormalized by Coulomb interactions. But its contributions can be strongly suppressed when $\ln (W/\Omega_D)$
is much larger than unity, a limit we will focus on in later discussions of applications. In this special limit, $\tilde{g}_{10}$ can be thought to be entirely determined by phonon interactions and Coulomb interactions play little role.

At last, surface interactions with $d_s=2$ follow the simple equations of
\begin{eqnarray}
\frac{\partial \tilde{g}_{1} }{\partial t_s} &=& ({d_s-1})  \tilde{g}_{1} + \frac{c_{sfp}}{\tilde{v}_F} \tilde{g}^2_{1}+ c_{sg} \alpha_{s} \tilde{g}_{1}  \nonumber\\
\frac{\partial \alpha_{s} }{\partial t_s }&=&c_s \alpha^2_s , \frac{\partial \tilde{v}_F}{\partial t}  = -c_{s} \alpha \tilde{v}_F
\nonumber\\
\tilde{g}_{1} (t_s=0 )&=& \tilde{g}_{10},\alpha_s (t_s=0)=\alpha_{s0}, \tilde{v}_F (t_s=0)=\tilde{v}_{F0}.
\nonumber \\
\label{RGEcphS}
\end{eqnarray}
where surface interactions at a {\em UV} scale $\Lambda_s=m$, $\tilde{g_{10}}$, $\alpha_{s0}$, have already been specified in Eq.\ref{match4}.
{\em We find that $c_{sg}=0$ as a consequence of absence of charge renormalization}. Because $\tilde{v}_{F}$ increases logarithmically in the IR limit, the quadratic term for the $\beta$ function of $\tilde{g}_{1}$ is less relevant in the infrared limit.
However, in the later discussions in Section IV, when Coulomb interactions are screened by an external metal, we can set $c_{sfp}/\tilde{v}_F$ to be one because in that case, there is no logarithmic growth in velocity $v_F$ and renormalization effects in $\tilde{v}_F$ can be neglected.

Solutions to Eq.\ref{RGEcphS} can be adopted as a guiding principle for discussions on possibilities of emergent surface topological superconductivity and surface Majorana particles induced by attractive interactions or spontaneous symmetry $U(1)$ breaking. They further provide valuable insight into the feasibility of making supersymmetric holographic matter via surface fermions.

But before we apply the principle of  {\em BBR} to understand these issues, we will further present a more quantitive implementation of {\em BBR} via a method of {\em dimension reduction}.
The principle of  naturally emerges as results integration over the degree of freedoms perpendicular to surfaces. For readers who are not interested in this special technology and a microscopic verification of  {\em BBR}, they can skip the next section
and go straight to Sec. IV for main predictions of {\em BBR} .

\section{Dimensional Reduction}

So far, we have introduced the general pedagogy of the interaction {\em bulk-boundary relation} based on the renormalization group methodology. In this section, we further establish the {\em BBR}  using the technique of {\em Dimensional Reduction(DR)}. Using this technique, one can also explicitly take into account the renormalization effect of bulk electrons on the surface field theory. In other words, the dimensional reduction technique will provide microscopic details of the bulk-boundary matching conditions stated at the beginning of the previous section.

In this technique, we let bosonic fields (photons or phonons) that mediate interactions between bulk or surface electrons be in the 3-dimensional bulk. Surface fermions have finite spatial extension along the z-axis, which is perpendicular to the topological insulator(TI) surface. Because the surface fermions have finite spreads into the interior of the bulk, the bulk bosons can mediate the interactions in the surface. This setup captures the essential features of interactions between boundary fermions and bulk bosonic fields. Later on, we integrate out the degrees of freedom perpendicular to the surface, thus replacing the 3-dimensional bulk theory with an effective 2-dimensional surface field theory.

In the first part, we discuss the dimensional reduction of the domain wall fermions mediated by the bulk boson fields at the tree level. Then, we take into account the polarization effect of bulk electrons on the respective bulk phonon and Coulomb fields. Thereafter, the dimensional reduction is then again performed on the renormalized action of the domain wall fermions to the lowest order in the bulk interactions strengths. Finally, we compare the result from dimensional reduction with that from the {\em RGE}, derived in the previous section using matching conditions at the bulk-boundary matching scale $m$. We find that the results obtained in two different approaches match, proving that {\em BBR} implemented through matching conditions at the mass scale provides an efficient way to take into account the effect of bulk polarization on surface interactions.

\subsection{Domain wall model}

 The domain wall model for surface fermions is defined as follows: consider that the TI surface is at $z = 0$ plane. The spatial profile of bulk mass is then,
\begin{eqnarray}
m(z) = \begin{cases} +m &  0 < z < +\infty\\ -m &  -\infty < z < 0
\end{cases} \label{mass}
\end{eqnarray}
Throughout our calculations, we ignore the edge effects on the surface physics, which means that the periodic boundary  conditions are applied on the 2D surface. By plugging this mass function to the massive Dirac equation and solving it, we obtain the wavefunction corresponding to domain wall fermions given by,
\begin{eqnarray}
      \Psi_{dw}(\textbf{r})_{k, \pm} &=& \frac{e^{i\textbf{k}_{\perp}.\textbf{r}}}{\sqrt{2}}\begin{bmatrix}
     1 \\ 0 \\ 0 \\ \pm e^{i\phi_{k_{\perp}}} 
    \end{bmatrix} \sqrt{\frac{m}{v_{F}}} e^{- \frac{m}{v_{F}} |z|}\nonumber \\ &=& \Psi_{s}(\textbf{r}_{\perp})_{k, \pm}\sqrt{\frac{m}{v_{F}}} e^{- \frac{m}{v_{F}} |z|}   \label{surfacestate}
\end{eqnarray}
Here $k_{\perp}$ = ($k_{x}$, $k_{y}$), $\textbf{r}_{\perp}$ = ($x$, $y$) and the angle $\phi_{k_{\perp}} = \text{tan}^{-1}\frac{k_{y}}{k_{x}}$. $\Psi_{s}(\textbf{r}_{\perp})$ is the 4-component spinor wavefunction confined to the 2D surface. Also, $v_{F}$ is the bare value of Fermi velocity. The dispersion relation for the states are given by $E_{\pm}(k) = \pm v_{F}|k_{\perp}|$. It is evident that the domain wall fermions are localized near the surface at $z = 0$, with the bulk mass term $m$ being the localization parameter. In the limit of $m \rightarrow \infty$, the domain wall state evolves itself into an ideal surface state with no tail end in the bulk. In the opposite limit where $m \rightarrow 0$, the bulk gap closes, and the surface states get hybridized with the bulk. In other words, there is no more bulk-boundary distinction. Thus the domain wall model for boundary electrons has the advantage that one could tune the localization of surface states as a function of the bulk mass term. Closing the bulk gap automatically merges the bulk and surface states.

Let us now write down the free-field theory of domain wall fermions and 2D surface fermions. For this purpose, we shall define a pseudo-spin space with the $4 \times 4$ matrices $\Xi_x$, $\Xi_y$ and $\Xi_z$ as the generators. These matrices have the definitions, 
\begin{eqnarray}
 \Xi_{x} &=& \frac{1}{2}\left(\tau_{x}\sigma_{x} - \tau_{y}\sigma_{y} \right), \,\, \Xi_{y} = \frac{1}{2}\left(\tau_{x}\sigma_{y} + \tau_{y}\sigma_{x} \right)\nonumber\\ \Xi_{z} &=& \frac{1}{2}\left(\tau_{z}\sigma_{0} + \tau_{0}\sigma_{z}\right)
 \end{eqnarray}
 where $\tau_{\alpha}$ $(\alpha = x,y,z)$ and $\sigma_{\alpha}$ $(\alpha = x,y,z)$ are Pauli matrices defined in the particle-hole and the spin-1/2 spaces respectively. It can be shown that the matrices $\Xi_x$, $\Xi_y$ and $\Xi_z$ satisfy the $SU(2)$ algebra given by, $\left[\Xi_{i}, \Xi_{j}\right] = 2 i \epsilon_{ijk}\Xi_{k}$. In this Hilbert space, the respective actions have the form,
\begin{eqnarray}
\mathcal{S}_{dwf} &=& \int d^{3}\textbf{r}dt\, \psi^{\dagger}_{dw}(\textbf{r}, t)\biggl[i\partial_{t} + i v_{F}\Xi_{x}\partial_{x}\nonumber\\  &+& i v_{F}\Xi_{y}\partial_{y}  + i v_{F}\partial_{z} + m(z)\biggr]\psi_{dw}(\textbf{r}, t) \nonumber \\ \mathcal{S}^{2D}_{surf} &=& \int d^{2}\textbf{r}dt\, \psi_{s}^{\dagger}(\textbf{r}, t)\biggl[i\partial_{t} + iv_{F}\Xi_{x}\partial_{x} + iv_{F}\Xi_{y}\partial_{y}\biggr]\psi_{s}(\textbf{r}, t)\nonumber\\ \label{surfacelagrangian}
\end{eqnarray}
 where $m(z)$ has the form given in Eq.\ref{mass}. Hence $\psi^{\dagger}_{dw}$ describe 4-component gapless domain wall fermions defined in the 3D bulk while $\psi^{\dagger}_{s}$  describes 4-component gapless surface fermions defined in the 2D surface. Thus the only difference between them is that the domain wall fermions have a finite spatial extension to the bulk which allow them to interact with the bulk Coulomb and the phonon field. 
\subsection{Dimensional reduction I}

In this sub-section, we write down the bare action that defines the interaction between the domain wall fermions. The repulsive interaction between domain wall fermions is mediated
by the Coulomb potential and the attractive
interaction by phonons, the quanta of lattice vibrations. This 3-dimensional theory is then mapped into an effective 2D surface theory by integrating out the degrees of freedom perpendicular to the surface. Thereafter, we look at the limit of the resulting surface theory which can lead to superconducting pairing and study the behavior of the two interactions at the tree level. Following this, we find out a relation between the dimensionless bulk and the surface coupling constants, thus verifying the bulk-boundary matching conditions stated in Eq.\ref{match4} at the tree level limit.

The dynamics of domain wall fermions subject to the phonon-mediated interaction and Coulomb repulsion are governed by the field theory:
\begin{eqnarray}
     \mathcal{S} &=& \mathcal{S}_{dwf} + \mathcal{S}_{e-ph, int} + \mathcal{S}_{cou, int} \nonumber \\
    \mathcal{S}_{e-ph, int} &=&  - G^{2}_{fp}\int d^{3}\textbf{r}\,d^{3}\textbf{r}'\,dt\,dt'\,\psi_{dw}^{\dagger}(\textbf{r}, t)\psi_{dw}(\textbf{r}, t)\nonumber\\ & &  K_{e-ph}(\textbf{r}' - \textbf{r}, t'-t)\psi_{dw}^{\dagger}(\textbf{r}', t')\psi_{dw}(\textbf{r}', t')\nonumber\\  \\ \mathcal{S}_{cou, int} &=& - \int d^{3}\textbf{r}\,d^{3}\textbf{r}'\,dt\, \psi_{dw}^{\dagger}(\textbf{r}, t)\psi_{dw}(\textbf{r}, t)\nonumber \\ & & \frac{e^{2}_{b}}{|\textbf{r} - \textbf{r}'|}\psi_{dw}^{\dagger}(\textbf{r}', t)\psi_{dw}(\textbf{r}', t) 
\end{eqnarray}
where $K_{e-ph}(\textbf{r}, t)$ is the  phonon propagator in the bulk of the topological insulator. In momentum space, it has the expression,
\begin{equation}
    K_{e-ph}(\textbf{q}, \omega) =   \frac{v^{2}_{p}q^{2}}{\omega^{2} - v^{2}_{p}q^{2} }. \label{phononprop3D}
\end{equation}
 Here  $v_{p}$ is the  phonon velocity. As discussed in the start of the section, the effective surface action is derived by integrating out the degrees of freedom perpendicular to the surface. That is, 
\begin{eqnarray}
\mathcal{S}^{2D} &=& \int^{\infty}_{-\infty} dz\, \left[ \mathcal{S}_{dwf} + \mathcal{S}_{e-ph, int} + \mathcal{S}_{cou, int} \right] \nonumber
\\ &=& \mathcal{S}^{2D}_{surf} + \mathcal{S}^{2D}_{e-ph, int} + \mathcal{S}^{2D}_{cou, int}
\end{eqnarray}
 The interaction term is now,
\begin{eqnarray}
    \mathcal{S}^{2D}_{e-ph, int} &=& - \int d^{2}\textbf{r}\,d^{2}\textbf{r}'\,dt\,dt'\,\psi_{s}^{\dagger}(\textbf{r}, t)\psi_{s}(\textbf{r}, t)\nonumber\\ & & K^{2D}_{e-ph}(\textbf{r}' - \textbf{r}, t'-t)\psi_{s}^{\dagger}(\textbf{r}', t')\psi_{s}(\textbf{r}', t')\nonumber \\ \label{dimredphon}\\  \mathcal{S}^{2D}_{cou, int} &=& - \int d^{2}\textbf{r}\,d^{2}\textbf{r}'\,dt\,\psi_{s}^{\dagger}(\textbf{r}, t)\psi_{s}(\textbf{r}, t)\nonumber\\ & &K^{2D}_{cou}(\textbf{r}' - \textbf{r})\psi_{s}^{\dagger}(\textbf{r}', t)\psi_{s}(\textbf{r}', t) \label{dimredcoulomb}
\end{eqnarray}
where $K^{2D}_{e-ph}(\textbf{r}, t)$ is now the effective phonon propagator on the 2D surface. Similarly, $K^{2D}_{cou}(\textbf{r})$ is the effective surface Coulomb propagator. In the momentum space, they have the form given by,
\begin{eqnarray}
    K^{2D}_{e-ph}(v_{p}\textbf{q}_{\perp}, \omega) &=&\nonumber \\ G^{2}_{fp} \int^{\infty}_{-\infty} & & \frac{dq_{z}}{2\pi} \left(\frac{M^{2}}{q_{z}^{2} + M^{2}}\right)^{2}\frac{ q_{\perp}^{2} + q^{2}_{z}}{\frac{\omega^{2}}{v^{2}_{p}} - q_{\perp}^{2} - q^{2}_{z}}\label{qzinteph} \\ K^{2D}_{cou}(\textbf{q}_{\perp}) &=& e^{2}_{b}\int^{\infty}_{-\infty} \frac{dq_{z}}{2\pi} \left(\frac{M^{2}}{q_{z}^{2} + M^{2}}\right)^{2} \frac{1}{q_{\perp}^{2} + q^{2}_{z}}\nonumber\\ \label{qzintcou}
\end{eqnarray}
where $M = \frac{2m}{v_{F}}$.

In the integral expression, the contribution from domain wall model $\left(\frac{M^{2}}{q_{z}^{2} + M^{2}}\right)^{2}$ is in-fact a regularization term that helps us get rid of ultraviolet divergences if any. If we take the limit of $M \rightarrow \infty$, the domain wall wavefunction converges into a strictly surface term. In the case of the Coulomb propagator, it is evident from Eq.\ref{qzintcou} that this integral is not UV divergent and hence is independent of $m$. Therefore, we could safely take the large $M$ limit and ignore the regularization term. 

But the integral involving bulk phonon propagator is UV divergent. Hence, we must use the regularization term to integrate out the $q_{z}$ momentum resulting in a linear dependence of $m$ . After taking all these into account, the effective surface propagator for phonon-mediated interaction and Coulomb interaction at the tree level turns out to be,
\begin{multline}
K^{2D}_{e-ph}(v_{p}\textbf{q}_{\perp}, \omega) = \\ \frac{G^{2}_{fp}M}{4\left(\omega^{2} - v^{2}_{p}q^{2}_{\perp} \right)}   \biggl[v^{2}_{p}q_{\perp}^{2}  - \frac{v^{2}_{p}M^{2}\omega^{2}}{ \left( Mv_{p} + \sqrt{v^{2}_{p}q_{\perp}^{2} - \omega^{2}}\right)^{2}}  \biggr]   \label{phGreenfunc} 
\end{multline}
\begin{gather} 
K^{2D}_{cou}(\textbf{q}_{\perp}) = \frac{e^{2}_{b}}{2q_{\perp}} \label{k(q)}.
\end{gather}
$K^{2D}_{cou}$, the 2D surface Coulomb propagator, turns out to be proportional to $\frac{1}{q_{\perp}}$ as expected, where $q_{\perp}$ is the in-plane momentum.

 As the main application of phonon-mediated interaction of $K^{2D}_{e-ph}$ is superconductivity, 
 we therefore restrict to the limit where the energy exchanged between the electrons in the scattering process is less than energies of most dominating phonons. 
 These exchange processes of phonons also lead to pairing near a Fermi surface\cite{abrikosov}. On taking the limit $\omega \ll v_{p}q_{\perp}, m$, the second term in the bracket in Eq.\ref{phGreenfunc} can be ignored in comparison to the first term, resulting in a rather simple form,
 \begin{eqnarray}
 K^{2D}_{e-ph}(v_{p}\textbf{q}_{\perp}, \omega) &\approx& \frac{G^{2}_{fp}m}{2v_{F}} \frac{v^{2}_{p}q_{\perp}^{2}}{\omega^{2} - v^{2}_{p}q^{2}_{\perp}}.    
 \end{eqnarray}
 
 Thus the momentum and frequency dependence of the 2D surface propagator are the same as its 3D bulk counterpart. It is well known that the phonon mediated interactions are mostly dominated by phonons near $v_p q_\perp \sim \Omega_D$,   
 In the low energy window where $\omega < m \ll \Omega_D$, we thus arrive at,
\begin{eqnarray}
    K^{2D}_{e-ph}(\omega, v_{p}\textbf{q}_{\perp}) &\sim& - \frac{G^{2}_{fp}m}{2v_{F}}\label{barephononpropIR} \\    K^{2D}_{cou}(\textbf{q}_{\perp}) &=& \frac{e^{2}_{b}}{2q_{\perp}}\label{barecoupropIR}
\end{eqnarray}

We find that, in this limit, the effective surface phonon propagator is short-ranged in space.
Upon comparing the surface action in Eqs.\ref{dimredphon},\ref{dimredcoulomb} with the interaction terms of the surface Hamiltonian $H_{surf}$ written down in Eq.\ref{Hfull}, one finds that the surface coupling constants can be related to their respective surface propagators as, 
\begin{eqnarray}
K^{2D}_{e-ph}(\omega, v_{p}\textbf{q}_{\perp}) =  g_{1}\,\,\,,\,\,\,\, q_{\perp}K^{2D}_{cou}(\textbf{q}_{\perp}) =  e^{2}_{s}.
\end{eqnarray}

Using this relation, one can connect the bulk and the surface coupling constants of the respective interactions. When represented in a dimensionless form using the scaling analysis, the bare values of the surface and the bulk coupling constants are found to be related in the following way at the tree-level limit,
\begin{eqnarray}
\tilde{g}_{10} = -\frac{\tilde{G}^{2}_{fp0}}{2} \left(\frac{m}{\Omega_{D}}\right)^{2}, \,\,\,\,\,  \tilde{e}^{2}_{s0} = \frac{\tilde{e}^{2}_{b0}}{2} \label{BBRtreelevel}
\end{eqnarray}
where $\tilde{g}_{10}= \tilde{g}(\lambda_{s}=m)$,  $\tilde{G}^{2}_{fp0}=\tilde{G}^{2}_{fp}(\Lambda_{b}=\Omega_{D})$  $\tilde{e}^{2}_{s0}=\tilde{e}^{2}_{s}(\Lambda_{s}=m)$ and $\tilde{e}^{2}_{b0}=\tilde{e}^{2}_{b}(\Lambda_{b}=W)$ are defined as the bare values of the dimensionless surface and the bulk coupling constants. Thus, up to the tree-level limit, we are able to verify the matching conditions between the respective surface and bulk interaction strengths stated in the previous section(Eq.\ref{match4}).

Upon dimensional reduction, one also finds that the low frequency limit of phonon-mediated interaction potential between surface fermions are attractive and short-ranged in space. Thus, if the Coulomb interaction is sufficiently suppressed in the long-wavelength scale, one can have a superconducting ground state on the topological insulator surface. In the subsequent sub-sections, we study the effect of bulk charge polarization on surface interactions using the diagrammatic method followed by Dimensional Reduction(DR).

\begin{figure}
\includegraphics[width=5cm]{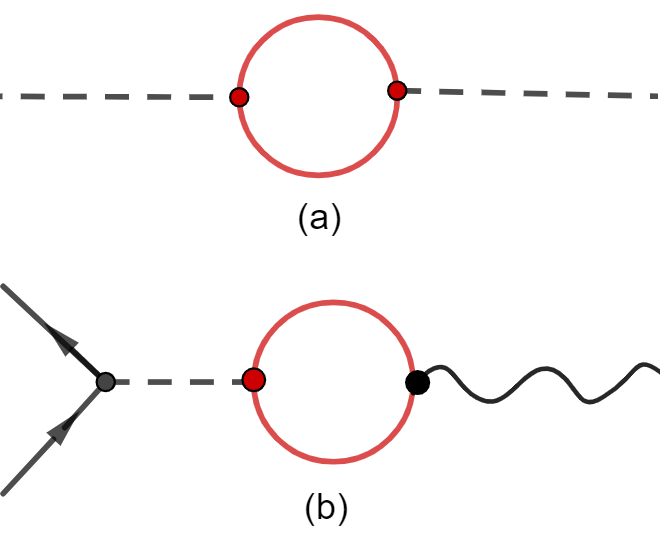}
\centering
\caption{\footnotesize Feynman diagrams for  (a) bulk charge polarization effects due to Coulomb interactions and (b) contributions of Coulomb interactions to renormalization of electron-phonon coupling constant. Dotted  and wiggly lines denote respectively Coulomb and phonon fields. Solid lines stand for bulk electron propagators.}
\label{diag_dimred}
\end{figure}

\subsection{Bulk renormalization effect}

The Coulomb and phonon fields could further nteract with massive Dirac fermions in the bulk, resulting in a series of electron-hole excitations and quantum renormalization effects. Here we study the one-loop renormalization effect due to bulk electrons on the boson fields. Later, we will perform dimensional reduction
of the domain wall fermions mediated by these dressed boson fields. The theory after dimensional reduction turns out to be an {\em UV}  theory of the surface, which includes the quantum corrections to the bulk-boundary matching conditions written in the previous section at the tree-level limit (Eq.\ref{BBRtreelevel}).

The field theory of bulk electrons of the topological state is governed by the action
\begin{equation}
    \mathcal{S}^{b}_{dirac} = \int d^{3}\textbf{r}\,dt\,  \psi^{\dagger}_{b}(\textbf{r}, t)\left( i\partial_{t} + i v_{F}\vec{\rho}.\vec{\nabla} - \eta m \right)\psi_{b}(\textbf{r}, t)
\end{equation}
where $\vec{\rho}$ and $\eta$ are $4\times 4$ matrices and are defined in terms of the Dirac gamma matrices as $\vec{\rho} = \gamma^{0}\vec{\gamma}$ and $\eta = \gamma^{0}$. $m$ is the mass gap of the bulk spectrum and $v_{F}$ is the Fermi velocity of the electrons. Note that, the Dirac field $\psi_{b}(r,t)$ defined here is a 4-component spinor and satisfies the anti-commutation relations.

The interaction between the bulk electrons is governed by the action,
\begin{equation}
    \mathcal{S}^{b}_{int} = \mathcal{S}^{b}_{cou, int} + \mathcal{S}^{b}_{e-ph,int}\nonumber
\end{equation}
where
\begin{eqnarray}
    \mathcal{S}^{b}_{cou, int} &=& - \int d^{3}\textbf{r}\,d^{3}\textbf{r}'\,dt\, \psi_{b}^{\dagger}(\textbf{r}, t)\psi_{b}(\textbf{r}, t)\nonumber\\ & & \frac{e_{b}^{2}}{|\textbf{r} - \textbf{r}'|}\psi_{b}^{\dagger}(\textbf{r}', t)\psi_{b}(\textbf{r}', t) \\ \mathcal{S}^{b}_{e-ph,int} &=& - G^{2}_{fp}\int d^{3}\textbf{r}\,d^{3}\textbf{r}'\,dt\,dt'\,\psi_{b}^{\dagger}(\textbf{r}, t)\psi_{b}(\textbf{r}, t)\nonumber\\ & &  K_{e-ph}(\textbf{r}' - \textbf{r}, t'-t)\psi_{b}^{\dagger}(\textbf{r}', t')\psi_{b}(\textbf{r}', t')\nonumber  \\
\end{eqnarray}

 Both the electric charge and the Fermi velocity can get renormalized in a 3-dimensional bulk state. As said before, in this section, we will present the standard one-loop renormalization. In this approximation, it is sufficient to use just the bare value of Fermi velocity. This is because the effects of Fermi self-energy renormalization will show up only in the higher order contributions to the bulk charge renormalization.

The charge polarization(see fig.\ref{diag_dimred}) term to the one loop order in electron bubble has the form\cite{Peshkin}, 
\begin{eqnarray}
     \Pi(\textbf{q}) &=& \frac{e^{2}_{b0}}{2\pi^{2}v_{F}}\int^{1}_{0} d\beta \, \beta \left(1 - \beta\right) \text{Log}\left[\frac{m^{2} + \beta(1 - \beta)v^{2}_{F}q^{2}}{m^{2} + \beta(1 - \beta)W^{2}}  \right]\nonumber\\ 
\end{eqnarray}
Note that $v_{F}$ here is the bare value of Fermi velocity.
The renormalization condition imposed is that the bulk renormalized charge must equal its bare counterpart at the bulk cut-off scale W, i.e $\Pi(v^{2}_{F}q^{2} = W^{2}) = 0$. Since we are more interested in the case where the energy scale W
is much greater in magnitude than the bulk mass m, the bulk polarization term takes the simple form,
\begin{eqnarray}
     \Pi(\textbf{q}) &\approx& \frac{e^{2}_{b0}}{2\pi^{2}v_{F}}\int^{1}_{0} d\beta \, \beta \left(1 - \beta\right) \text{Log}\left[\frac{m^{2} + \beta(1 - \beta)v^{2}_{F}q^{2}}{\beta(1 - \beta)W^{2}}  \right]\nonumber\\ \label{selfenergy_cou}
\end{eqnarray}
As a result of the polarization, the bulk electric charge gets renormalized and hence becomes a function of spatial momentum(see fig. \ref{diag_dimred}(a)), given by, 
\begin{eqnarray}
e^{2}_{b}\left(\textbf{q}\right) = \frac{e^{2}_{b0}}{1 - \Pi(\textbf{q})}
\end{eqnarray}
where $e^{2}_{b0}$ is the bare charge of the bulk electrons at cut-off scale $W$.

 We have seen in the previous section that the Coulomb field can renormalize the electron-phonon interaction vertex. The corresponding one-loop Feynman diagram is shown in fig.\ref{diag_dimred}(b). The renormalized bulk electron-phonon coupling constant has the form,   
\begin{eqnarray}
G_{fp}\left(\textbf{q}\right) = \frac{G_{fp0}}{1 - \Pi^{eph}(\textbf{q})}
\end{eqnarray}
$\Pi^{eph}(\textbf{q})$ has the same loop structure as the charge polarization but with the UV cut off set at Debye frequency, $\Omega_{D}$, and $G_{fp0}$ is the bare value of coupling constant at the cut-off scale. That is,
\begin{eqnarray}
  \Pi^{eph}(\textbf{q}) &=& \frac{e^{2}_{b}(\Omega_{D})}{2\pi^{2}v_{F}}\int^{1}_{0} d\beta \, \beta \left(1 - \beta\right)\nonumber\\ & & \text{Log}\left[\frac{m^{2} + \beta(1 - \beta)v^{2}_{F}q^{2}}{ \beta(1 - \beta)\Omega_{D}^{2}}  \right] \label{selfenergy_e-ph}
\end{eqnarray}
In both $\Pi({\bf q})$ and $\Pi^{eph}({\bf q})$, there is a standard logarithmic dependence on $m$ in the limit of zero momentum ${\bf q}=0$. 

\subsection{Dimensional Reduction II}

Now we shall come back to talk about the domain wall fermions of the surface. We have learned  that the polarization effect of bulk fermions renormalize the respective interaction strengths. These dressed boson fields could mediate interactions between domain wall fermions, as shown in the schematic diagram in fig.\ref{bulkboundary}. Following the same procedure as was done in the previous sub-section, we perform dimensional reduction of the renormalized theory, to arrive at the effective surface theory. The resulting surface field theory has the form,
\begin{eqnarray}
         \mathcal{S}^{2D}_{e-ph, int} &=&  -\int d^{2}\textbf{r}\,d^{2}\textbf{r}'\,dt\,dt'\,\psi_{s}^{\dagger}(\textbf{r}_{\perp}, t)\psi_{s}(\textbf{r}, t)\nonumber\\ & &  \mathcal{K}^{2D}_{e-ph}(\textbf{r}' - \textbf{r}, t'-t)\psi_{s}^{\dagger}(\textbf{r}', t')\psi_{s}(\textbf{r}', t')\nonumber\\ \mathcal{S}^{2D}_{cou, int} &=&  -\int d^{2}\textbf{r}\,d^{2}\textbf{r}'\,dt\, \psi_{s}^{\dagger}(\textbf{r}, t)\psi_{s}(\textbf{r}, t)\nonumber \\ & & \mathcal{K}^{2D}_{cou}\left(\textbf{r}' - \textbf{r} \right)\psi_{s}^{\dagger}(\textbf{r}', t)\psi_{s}(\textbf{r}', t) \nonumber\\
\end{eqnarray}
where $\mathcal{K}^{2D}_{e-ph}(\textbf{r}_{\perp}, t)$ and $\mathcal{K}^{2D}_{cou}(\textbf{r}_{\perp})$ differ from the bare surface propagators defined in the subsection B in such a way that these are obtained by performing dimensional reduction on the dressed bulk propagators, that take into account the bulk renormalization effect[defined in Eqs. \ref{selfenergy_cou}, \ref{selfenergy_e-ph}].
\begin{eqnarray}
\mathcal{K}^{2D}_{cou}(\textbf{q}_{\perp}) &=& \int^{\infty}_{-\infty} \frac{dq_{z}}{2\pi} \frac{e^{2}_{b0}}{1 - \Pi(\textbf{q})} \frac{1}{q_{\perp}^{2} + q^{2}_{z}} \\  \mathcal{K}^{2D}_{e-ph}(\textbf{q}_{\perp}, \omega) &=& \int^{\infty}_{-\infty} \frac{dq_{z}}{2\pi} \left(\frac{M^{2}}{q_{z}^{2} + M^{2}}\right)^{2}\nonumber\\ & &\frac{G^{2}_{fp0}}{\left(1 - \Pi^{eph}(\textbf{q}) \right)^{2}}\frac{ v^{2}_{p}\left(q_{\perp}^{2} + q^{2}_{z}\right)}{\omega^{2} - v^{2}_{p}\left(q_{\perp}^{2} + q^{2}_{z}\right)}\nonumber \\
\label{dressed_prop}
\end{eqnarray}
Computing this integral is quite tricky. So, let us look for expansions that help us get the lowest order corrections to bare result. In this case, we perform a perturbative expansion of the denominator in powers of the bare bulk coupling constant $\frac{e^{2}_{b0}}{v_{F}}$. This result in, 
\begin{widetext}
\begin{eqnarray}
\mathcal{K}^{2D}_{cou}(\textbf{q}_{\perp}) &=& e^{2}_{b0}\int^{\infty}_{-\infty} \frac{dq_{z}}{2\pi} \left(\frac{M^{2}}{q_{z}^{2} + M^{2}}\right)^{2}\frac{1}{q_{\perp}^{2} + q^{2}_{z}} \left[1 + \Pi(\textbf{q}) + ....\right] = \mathcal{K}_{cou}^{2D (0)}(\textbf{q}_{\perp}) + \mathcal{K}_{cou}^{2D (1)}(\textbf{q}_{\perp}) + .....\nonumber \\ \label{effsurfacepropCou}
\\
\mathcal{K}^{2D}_{e-ph}(\textbf{q}_{\perp}, \omega) &=& G^{2}_{fp0}\int^{\infty}_{-\infty} \frac{dq_{z}}{2\pi} \left(\frac{M^{2}}{q_{z}^{2} + M^{2}}\right)^{2}\frac{ v^{2}_{p}\left(q_{\perp}^{2} + q^{2}_{z}\right)}{\omega^{2} - v^{2}_{p}\left(q_{\perp}^{2} + q^{2}_{z}\right)} \left[1 + 2\Pi^{eph}( \textbf{q}) + .... \right]  = \mathcal{K}_{e-ph}^{2D(0)}(\textbf{q}_{\perp}) + \mathcal{K}_{e-ph}^{2D(1)}(\textbf{q}_{\perp}) + ....\nonumber\\\label{effsurfacepropeph}
\end{eqnarray}

\end{widetext}
where the zeroth order terms $\mathcal{K}_{e-ph}^{2D(0)}(\textbf{q}_{\perp})$, $\mathcal{K}_{cou}^{2D(0)}(\textbf{q}_{\perp})$ are the same as $K_{e-ph}^{2D}(\textbf{q}_{\perp})$ and $K_{cou}^{2D(0)}(\textbf{q}_{\perp})$ derived in the previous sub-section and their exact expressions were given in Eqs. \ref{phGreenfunc}, \ref{k(q)} respectively.

\paragraph*{First order correction}
Here we shall solve for $\mathcal{K}^{2D(1)}_{cou}(\textbf{q}_{\perp})$ and $\mathcal{K}^{2D(1)}_{e-ph}(\textbf{q}_{\perp}, \omega)$, the lowest order quantum corrections to the bare propagators corresponding to the Coulomb and phonon-mediated interaction respectively. 
Even though the complete solution of the momentum integral is cumbersome, we limit ourselves to the case where the magnitude of the in-plane momentum is close to the bulk mass gap and the phonon frequency set to small, i.e. the static limit. The theory after performing dimensional reduction will manifest itself as the {\em UV} theory of the surface. Since the mass gap $m$ is the UV cut-off of the surface theory, it is justified to take the in-plane momentum to be of the order of bulk mass. The static limit suggesting pairing interactions that we will be interested in. In other words, the first order quantum correction to the surface propagators at the mass scale and the static limit manifest as the first-order correction to the bare values of their respective surface coupling constants,
\begin{eqnarray}
q_{\perp}\mathcal{K}^{2D(1)}_{cou}(\textbf{q}_{\perp})|_{v_{F}q_{\perp} \approx m} &=& e^{2(1)}_{s0}\nonumber \\ \mathcal{K}^{2D(1)}_{e-ph}(\textbf{q}_{\perp}, \omega)|_{v_{F}q_{\perp} \approx m, \omega \approx 0} &=& g^{(1)}_{10} 
\end{eqnarray}

After integrating out $q_{z}$, one finds the exact analytical form of $e^{2(1)}_{s0}$ and $g^{(1)}_{10}$ as,
\begin{eqnarray}
e^{2(1)}_{s0} &=& \frac{e^{2}_{b0}}{2} \frac{e^{2}_{b0}}{6\pi^{2}v_{F}}\left[ \log \frac{m}{W} + 1.2611\right] \label{dressedcoufirstorderIR} \\   g^{(1)}_{10} &=&  - \frac{G^{2}_{fp0}m}{2v_{F}} \frac{e^{2}_{b}(\Omega_{D})}{3\pi^{2}v_{F}}\biggl[\log \frac{m}{\Omega_{D}} + 1.06874\biggr]\label{dressedphonfirstorderIR}
\end{eqnarray}
Details of the calculation is shown in Appendix \ref{Appendix}.

\subsection{Comparison with RG results}
In this section so far, we have used the technique of dimensional reduction to project the renormalized bulk theory onto the surface. Thereafter, we have derived the lowest order quantum correction due to the bulk electron polarization to the surface field theory. In this sub-section, we will look at the surface coupling constants derived in the previous section using RG techniques and compare them with the results from dimensional reduction. At the beginning of section II, we have stated that the effect of bulk interaction can be taken into account for the surface theory by demanding that the bulk and the boundary coupling constants must satisfy the matching condition at the energy scale at which the boundary connects with the bulk, which in this case is the bulk mass $'m'$. Our objective of introducing the technique of dimensional reduction was to provide a microscopic details of these bulk-boundary matching conditions in the case of a 3D topological state.

The bulk-boundary matching condition is written down in Eq.\ref{match4}. Expanding the relation to the first order in the bulk Coulomb interaction strength $\frac{e^{2}_{b0}}{v_{F}}$, we arrive at 
\begin{eqnarray}
\tilde{e}^{2}_{s0} (e_{b0}, \frac{m}{W})&=& \tilde{e}^{2}_{b0}\left[1 + \frac{\tilde{e}_{b0}^{2}}{6\pi^{2}v_{F}} \ln \frac{m}{W}  + ... \right] \nonumber \\
\tilde{g}_{10}(\tilde{G}_{fp0}, \frac{m}{\Omega_D}) &=&-\tilde{G}^2_{fp0}\left(\frac{m}{\Omega_{D}} \right)^{2}\left[1 + \frac{\tilde{e}_{b}^{2}\left(\Omega_{D} \right)}{3\pi^{2}v_{F}} \ln \frac{m}{\Omega_{D}} + .. \right]\nonumber \\
\end{eqnarray}
On the other hand, from Eqs.\ref{BBRtreelevel}, \ref{dressedcoufirstorderIR} and \ref{dressedphonfirstorderIR}, the dimensionless surface coupling constants derived via the Dimensional Reduction technique to the lowest order in quantum corrections are ,
\begin{eqnarray}
\tilde{e}^{2}_{s0} &=& \tilde{e}^{2(0)}_{s0} + \tilde{e}^{2(1)}_{s0} + .....\nonumber \\ &=& \frac{\tilde{e}^{2}_{b0}}{2}  \biggl[1 + \frac{\tilde{e}^{2}_{b0}}{6\pi^{2}v_{F}}\biggl( \ln \frac{m}{W} + 1.2611\biggr) + ... \biggr]\nonumber \\
\tilde{g}_{10} &=& \tilde{g}^{(0)}_{10} + \tilde{g}^{(1)}_{10} + .......\nonumber \\ &=&- \frac{\tilde{G}^{2}_{fp0}}{2} \left(\frac{m}{\Omega_{D}}\right)^{2} \biggl[1  + \frac{\tilde{e}^{2}_{b}(\Omega_{D})}{3\pi^{2}v_{F}}\biggl(\ln \frac{m}{\Omega_{D}}\nonumber \\ &+& 1.06874\biggr) + ....\biggr]
\end{eqnarray}
where the dimensionless form of the bulk coupling constants at their respective UV cut-off scales has the definitions, $\tilde{e}_{b0} = e_{b0}$ and $\tilde{G}_{fp0} =G_{fp0} \left(\frac{\Omega_{D}}{v_{F}}\right)^{(D_b-1)/2}$ respectively, with $D_b=3$ for bulk. Similarly, for the bare surface coupling constants, $\tilde{e}_{s0} = e_{s0}$ and $\tilde{g}_{10} = g_{10}\left(\frac{m}{v_{F}}\right)^{d_{s} - 1}$ with $d_{s} = 2$ for the surfaces.

 Thus we find that, except for a numerical factor, the dimensional reduction technique nicely recreates the RG results obtained by imposing the matching conditions at the bulk-boundary matching scale $m$. Thus, we show that the bulk-boundary correspondence indeed naturally emerges as a result of integrating over the degrees of freedom perpendicular to the surfaces.

 To conclude, in this section we utilized the technology of dimensional reduction to show that the bulk-boundary matching conditions at the matching scale is an efficient way to take into account the bulk renormalization effect on the surface field theory. It should be noted that since the bulk mass $m$ is the UV cut-off scale for surface fermions, renormalization effects due to surface fermions come into the picture once the energy scale is further brought down towards an infrared regime of the surface theory. The RG flow of the surface coupling constants as a result of surface fermion renormalization has been written down in Eq.\ref{RGEcphS} in the previous section.

\section{Applications} 

For a three dimensional topological bulk, a natural limit for us to focus on is when $W$, the bulk bandwidth that can be associated with the {\em UV} scale of Coulomb interaction is much higher
than either the surface "bandwidth" or bandgap $m$ or  $\Omega_D$, the Debye frequency that sets the bandwidth of phonons. 
As seen in previous sections, this can result in substantial suppression of long range Coulomb interactions at lower energy windows of $m$ or $\Omega_D$ due to strong renormalization effects.

However, even if phonon-mediated interactions dominate at the scale of $\Omega_D$ due to strong suppression of long-range Coulomb interactions,
long range Coulomb interactions are represented by marginal surface operators and can still become dominating again at asymptotic regimes where $\Lambda_s \ll m, \Omega_D$. 
This hinders the emergence of surface topological superconductivity and other related interaction phenomena. The path suggested below involves two key ingredients.

\begin{figure}
\includegraphics[width=6cm]{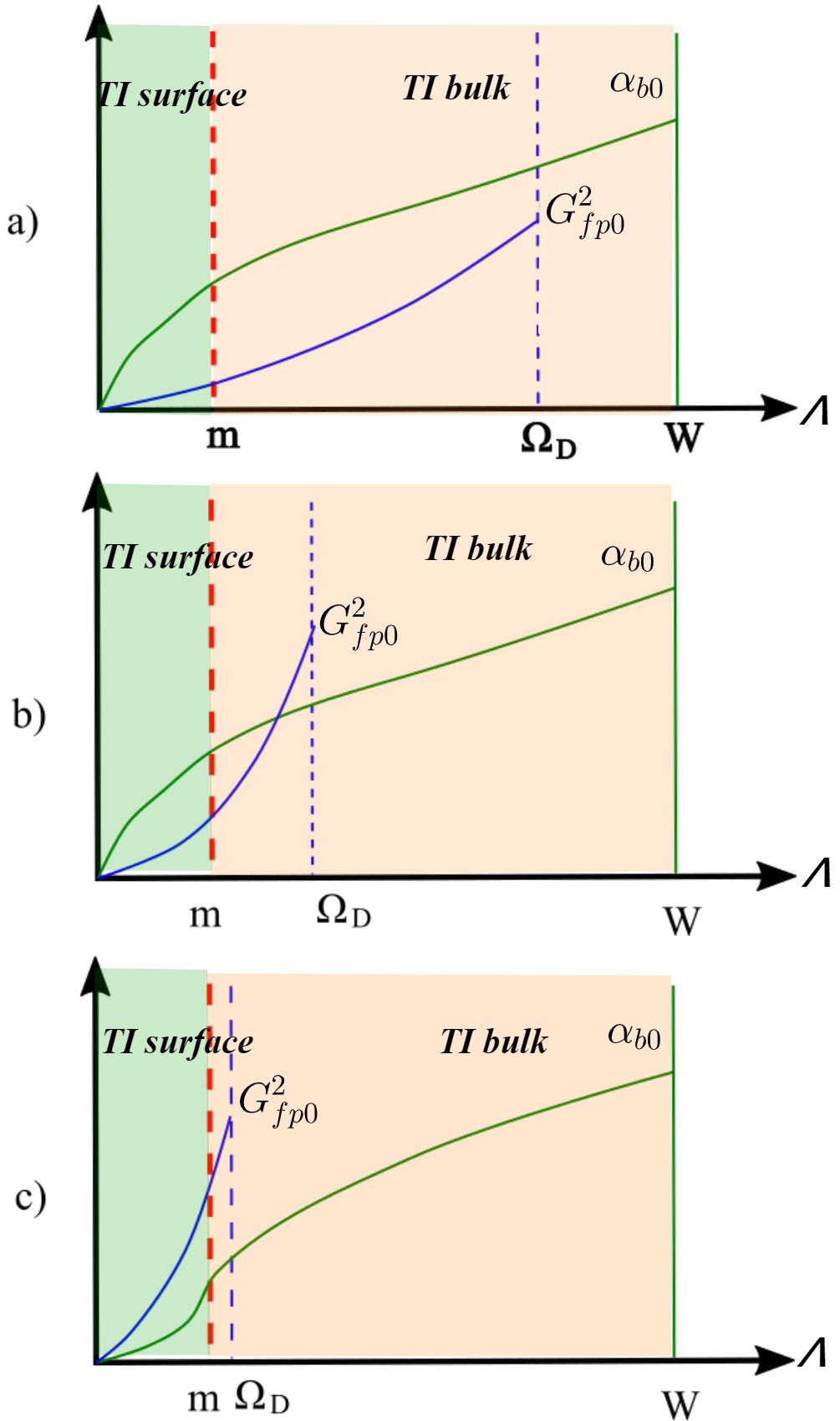}
\centering
\caption{ \footnotesize Schematic of renormalized Coulomb interactions (Green curve) and phonon-mediated interactions (Blue one)  vs the running scale $\Lambda$ in the renormalization equations.
In a)-c), we characterize the surface bandwidth with a mass gap $m$, and bulk phonon bandwidth $\Omega_D$ and
bulk fermion band width $W$. a) the limit of $W \sim  \Omega_D  \gg m$; b) the limit of $W \gg \Omega_D, m$ and c) the limit of $W \gg \Omega_D \sim m$.
In c), Coulomb interactions further turn into dipole interactions when $\Lambda \ll m$  because of a screening metal. Here $\alpha_{b0}$ is the bare Coulomb coupling constant (see Eq.\ref{RGEc}) and $\tilde{G}_{fp0}$ the dimensionless electron-phonon one (see Eq.14).
 }
\label{RGE BB.png}
\end{figure}
\begin{figure}
\includegraphics[width=8cm]{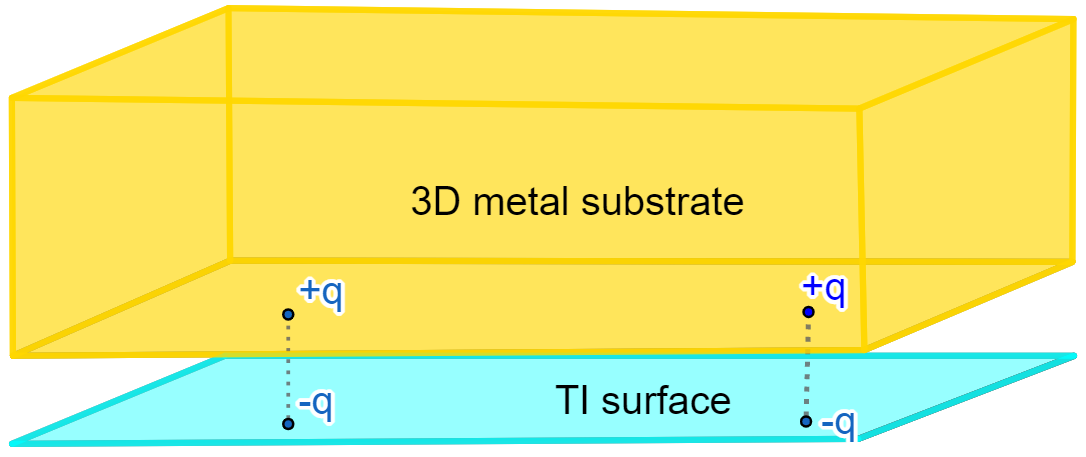}
\centering
\caption{ \footnotesize An illustrative picture representing the effect of a metal substrate on top of a topological insulator(TI) surface. 
}
\label{metal}
\end{figure}


i) A route to avoid this issue of long-range Coulomb interaction is to further add a metal on top of surfaces to effectively reduce long-range Coulomb interactions into a dipole-dipole one (see fig.\ref{metal}). A Similar idea was proposed in a previous study\cite{Ponte14}.
Surface dipole-dipole interactions are represented by an irrelevant operator that scales down much faster than phonon-mediated interactions.
So at least asymptotically, surfaces can then interact mainly via phonon-mediated interactions. One shall anticipate that there shall be spontaneous $U(1)$ symmetry breaking leading to surface topological
superconductivity that supports emergent quantum Majorana particles\cite{QiZhang,kitaev}.

ii) In the limit we are interested in, the Debye frequency $\Omega_D$ is larger or much larger than the mass gap $m$, i.e. $\Omega_D  \geq m$. 
Once identifying surface interactions at the scale $m$ via matching conditions in  {\em BBR}, 
one can extract the bare interaction constant ${g}_1(\Lambda_s=\Omega_D)$ (dimensionful) or $\tilde{g}_1(\Omega_D)$ (dimensionless one) via tracking the {\em UV} flow of {\em RGE}s from the intermediate scale $m$ to $\Omega_D$.

If we focus on the weakly interacting limit, the procedure mentioned above is particularly straightforward.
The {\em bare} interaction constant is simply related to $\tilde{g}_1(m)$ via $g_1(\Omega_D) =\tilde{g}_1(m) /m^{d_s-1}$.
 $\tilde{g}_1(\Omega_D)$, the dimensionless phonon mediated interactions at $\Lambda_s=\Omega_D$ are then linearly proportional to the interactions defined at matching scale $\Lambda_s=m$ that had been obtained via the {\em BBR}. 
This relation also follows the {\em RGE}s in Eq.\ref{RGEcphSD} obtained for surface fermions but in the limit when $\tilde{g}_1^2$ term is muted to neglect weak surface renormalization effects, i.e. following a tree level scaling.

 Therefore, one finds that

\begin{eqnarray}
\tilde{g}_1 (\Lambda_s=\Omega_D) &=&\tilde{g}_{1}( \Lambda_s=m) (\frac{\Omega_D}{m})^{d_s-1}
\nonumber \\
&=&-{\tilde{G}^2_{fp0}}(\frac{m}{\Omega_D})^{D_b-d_s} \nonumber \\
\end{eqnarray}
where $d_s=2$ is the dimension of surfaces and $D_b=3$ is the bulk dimension.

For the purpose of realizing surface topological superconductivity and possible supersymmetry surface holographic matter, one can further optimize phonon mediated surface interactions
by choosing a topological bulk so that

\begin{eqnarray}
m \approx \Omega_D \ll W.
\end{eqnarray}
By doing so, one can achieve an optimal phonon mediated interaction for given bulk fermion-phonon interactions, i.e
 
 \begin{eqnarray}
 \tilde{g}_{1} (\Lambda_s=\Omega_D) = \tilde{g}_{10}(\tilde{G}_{fp0}, \frac{m}{\Omega_D}=1)& =&- {\tilde{G}^2_{fp0}},  \nonumber \\
 \label{match5}
 \end{eqnarray}
 Compared with the general result in Eq.\ref{match4}, $\tilde{g}_{10}$ is equal to $\tilde{G}^2_{fp0}$ in this limit,
 where all other suppression factors due to $\frac{m}{\Omega_D} \leq 1$ are simply reduced to unity, a maximum value.
 So in the limit $\Omega_D \geq m$ and for a given {\em dimensionless} electron-phonon interaction constant $\tilde{G}^2_{fp0}$, the surface attractive interaction reaches a maximum when $m \sim \Omega_D$.     
Intuitively, one can relate the matching of surface bandwidth $m$ with Debye frequency $\Omega_D$ to one kind of {\em dynamic resonance}. Below we will focus on this limit only.

Construction based on point i) and ii) leads to the following effective field theories for interacting surfaces

\begin{eqnarray}
{\mathcal H}_{surf} &=&  v_{F}\int d^2{\bf r} \psi_s^\dagger {\bf s}\cdot {\bf p}  \psi_s \nonumber \\
&+& g_1
\int d^{2}{\bf r} \psi_s^\dagger ({\bf r}) \psi_s({\bf r}) \psi_s^\dagger({\bf r}) \psi_s({\bf r}) \nonumber \\
&+& p_{s}^2
\int d^{2}{\bf r} d^2{\bf r'} \psi_s^\dagger ({\bf r}) \psi_s({\bf r})  \frac{1}{|{\bf r} -{\bf r}'|^3} \psi_s^\dagger({\bf r}') \psi_s({\bf r}').\nonumber \\
\end{eqnarray}
Under scale transformation, the dimensionless coupling constants, $\tilde{g}_1=g_1 \Lambda^{d_s-1}$ and $\tilde{p}^2_s=e^2_s (\Lambda_s/m)^2$,  transform according to the following one-loop renormalization group
equations ($t_s =\ln \Lambda_s/m$),

\begin{eqnarray}
\frac{\partial \tilde{g}_{1} }{\partial t_s} &=& ({d_s-1})  \tilde{g}_{1} + \tilde{g}^2_{1}  \nonumber\\
\frac{\partial \tilde{p}^2_{s} }{\partial t_s }&=& 2 \tilde{p}^2_s \nonumber \\
\tilde{g}_{1} (t_s=0 )&=& \tilde{g}_{10},\tilde{p}^2_s (t_s=0)=\tilde{e}^2_{s0}.
\label{RGEcphSD}
\end{eqnarray}
Below we are going to discuss implications of the principle of {\em BBR} by utilizing solutions to Eq.\ref{RGEcphSD}

\subsection{Surface Topological Superconductivity and Majorana Fermions}

If surfaces are dominated by phonon-mediated interactions, topological superconductivity can naturally emerge in the presence of finite chemical potential $\mu$ via conventional spontaneous $U(1)$ symmetry breaking.
This can happen even in weakly interacting limits (although practically finite temperature measurements can put certain low bounds on interactions.)
If surfaces are weakly interacting, then Eq.\ref{RGEcphS} leads to the following simple solution  for $\Lambda_s \leq m$,

\begin{eqnarray}
\tilde{g}_{1} (\Lambda_s)= \tilde{g}_{10} (\frac{\Lambda_s}{m}), \mbox{	or	} \tilde{p}_s (\Lambda_s)=\tilde{e}^2_{s0} (\frac{\Lambda_s}{m})^2.
\end{eqnarray}
It is evident that if we choose to work with crystals with $m \sim \Omega_D \ll W$, a sufficient condition to realize surface topological superconductivity via interactions is

\begin{eqnarray}
\tilde{g}_{10} > \alpha_{s0}, \mbox{	or	} 
{\tilde{G}^2_{fp0}} > \frac{\alpha_{b0}}{1+c_b \alpha_{b0} \ln \frac{W}{m}} 
\end{eqnarray} 
where the second inequality in terms of bulk interaction parameters is obtained by applying Eq.\ref{match4}.

The effective field theory that describe emergent topological superconductivity phenomena and majorana field hosted in surfaces can be

\begin{multline}
{\mathcal H}_{surf}-\mu N  = \\ \int d^2{\bf r} \chi_s^T \biggl(-i\mathbb{I} \otimes {\bf e}_{y} \times {\bf s}\cdot {\bf \nabla} - \tau_z \otimes s_y\Delta -  \mu \tau_y \otimes \mathbb{I}  \biggr)\chi_s \\ + g_M
\int d^{2}{\bf r} \chi_s^T ({\bf r}) \tau_y \otimes \mathbb{I} \chi_s({\bf r}) \chi^T_s({\bf r}) \tau_y \otimes\mathbb{I}  \chi_s({\bf r})
\label{majorana}
\end{multline}
Here $\chi_s^T=(\chi_{1\uparrow},\chi_{1\downarrow},\chi_{2\uparrow},\chi_{2\downarrow})$ are four-component spinfull majorana fermion fields that forms a natural representation of two-component
surface fermions. In Eq.\ref{majorana}, we assume that the surface is oriented normal to the y-direction. 
$\tau_\alpha$, $\alpha=x,y,z$ act on a subspace spanned by subscript indices $1,2$ and $s_\alpha$, $\alpha=x,y,z$ act on spin indices $\uparrow,\downarrow$.
And $\chi_s$-fields satisfy the standard algebra of real-fermions

\begin{eqnarray}
\chi^\dagger_{s}({\bf r})=\chi^T_s({\bf r}),
\{ \chi_s({\bf r}), \chi_s({\bf r}') \}=\mathbb{I} \otimes \mathbb{I}  \delta({\bf r}-{\bf r}')
\end{eqnarray}
where two unity matrices are defined for two subspaces introduced above respectively.
In presenting Eq.\ref{majorana}, we have restricted ourselves to density-density interactions although it is possible to include other interactions
such as spin-density interactions. Detailed dynamics induced by other interactions are not quite relevant to our current discussions of surface topological superconductivity and will be presented in a separate article in preparation.

Topological surface superconductivity belongs to a class of symmetry-protected states\cite{Ludwig, QiZhang,kitaev} and it is stable with respect to weak interactions represented by $g_M$. Finite pairing amplitude $\Delta$ in Eq.\ref{majorana} induced by any weak interactions
at a finite chemical potential, $\mu$, is driven by the standard Cooper instability as in metals.
On topological surfaces,
this not only breaks $U(1)$ symmetry spontaneously but further results in emergent relativistic Majorana fields. On the other hand, if $\Delta$ is taken to be zero, Eq.\ref{majorana} has an additional $U(1)$ invariance under a rotation around $\tau_y$ axis. We identify it as a global $U(1)$ symmetry of surface complex fermions and the theory restores U(1) symmetry as anticipated.

\subsection{Supersymmetry Surface Holographic Matter} 

It is perhaps more interesting to consider the limit of strongly interacting surfaces. This limit is associated with a phenomenon of emergent supersymmetry suggested and studied before in a few different condensed matter systems including
topological surfaces\cite{Lee07,Grover14,Ponte14,Jian15,Li17,Li18,Jian17}.  Supersymmetry states are conformal invariant and can be easily identified with scale-invariant solutions of effective surface field theories we are examining.
Indeed, Eq.\ref{RGEcphSD} has a strong coupling fixed point at\cite{Sachdev}

\begin{eqnarray}
\tilde{g}_1^*= -({d_s-1}).
\label{fp}
\end{eqnarray}
We identify it as a quantum critical point (QCP) for transitions from a weakly interacting gapless topological surface to a gapped topological surface superconductor due to spontaneously $U(1)$ symmetry breaking at zero chemical potential or $\mu=0$. 
The effective surface theories for such a QCP can be obtained from Eq.\ref{majorana} by setting $\Delta=\mu=0$, i.e.,
\begin{eqnarray}
{\mathcal H}_{surf} &=&  \int d^2{\bf r} \chi_s^T \mathbb{I} \otimes {\bf e}_y \times {\bf s}\cdot \frac{{\bf \nabla}}{i}   \chi_s \nonumber \\
&+& g_M
\int d^{2}{\bf r}  [ \chi_s^T ({\bf r}) \tau_x \otimes {s_y} \chi_s({\bf r}) \chi^T_s({\bf r}) \tau_x \otimes {s_y}  \chi_s({\bf r}) \nonumber \\
&+& \chi_s^T ({\bf r}) \tau_z \otimes {s_y} \chi_s({\bf r}) \chi^T_s({\bf r}) \tau_z \otimes {s_y}   \chi_s({\bf r}) ].
\label{majorana2}
\end{eqnarray}
Here we have used a representation for 4-fermion operators that is most convenient for discussions on symmetries.
In this representation, surfaces are invariant under a $U(1)$ rotation generated around $\tau_y$, i.e. $U(\theta) =e^{i\theta\tau_y/2}$. 

As stated in the previous subsection, this is an emergent $U(1)$ symmetry that can be identified with charge conservation.
One can show explicitly this describes surface complex fermions near a supersymmetry 
 QCP. Eq.\ref{fp} implies that in the vicinity of strong coupling,
 
 \begin{eqnarray}
 {\tilde{G}^2_{fp0}} =(d_s-1),
 \end{eqnarray}
 one can anticipate surfaces become a conformal field theory state with supersymmetry. 
 
 \section{Conclusions}

In conclusion, we have put forward in this article a general bulk-boundary relation of interactions and have  illustrated its application in a few concrete limits.
We have also investigated this principle in a diagrammatic-based dimensional reduction approach. This general principle has been employed to understand the conditions for spontaneous symmetry breaking to occur on topological surfaces.
We further discuss a potential route to supersymmetric matter via strong electron-phonon interactions and additional metallic screening of long-range Coulomb interactions.
It is possible to apply the principle to other approaches to topological surface symmetry breaking phenomena. We plan to explore this direction in near future. 

This project is funded by NSERC(Canada) discovery grant under RGPIN-2020-07070. One of us (F.Z.) wants to thank discussions with Ian Affleck and Fan Yang.
\appendix
\section{}
\label{Appendix}
Here we describe the step by step solution to the $q_{z}$ momentum integral in Eq.\ref{effsurfacepropCou}, \ref{effsurfacepropeph} which contributes to the quantum correction to the effective surface Coulomb and phonon propagators respectively. Let us solve for the correction term to the Coulomb propagator first.   

\subsection*{Evaluation of $\mathcal{K}_{cou}^{2D(1)}(\textbf{q}_{\perp})$}

It was already mentioned in the main text that, in the case of Coulomb propagator, the momentum integral is UV convergent. Therefore the regularization term $\left(\frac{M^{2}}{q_{z}^{2} + M^{2}}\right)^{2}$ can be safely ignored by taking the limit of $M \rightarrow \infty$ ( where $M = \frac{2m}{v_{F}}$). This results, in the first order correction term to have the form given by,  
\begin{widetext}
\begin{eqnarray}
\tilde{K}_{cou}^{2D(1)}(\textbf{q}_{\perp}) &=& \frac{e^{4}_{b0}}{2\pi^{2}v_{F}}  \int \frac{dq_{z}}{2\pi} \frac{1}{q_{\perp}^{2} + q^{2}_{z}} \Pi\left(q_{\perp}, q_{z}\right) \\ &=& \frac{e^{4}_{b0}}{2\pi^{2}v_{F}}  \int \frac{dq_{z}}{2\pi} \frac{1}{q_{\perp}^{2} + q^{2}_{z}}\int^{1}_{0} d\beta \,\beta \left(1 - \beta\right) \text{Log}\left[\frac{m^{2}}{\beta(1 - \beta)W^{2}} + \frac{v^{2}_{F}\left(q_{\perp}^{2} + q^{2}_{z}\right)}{W^{2}}\right]
\end{eqnarray}
\end{widetext}
\begin{figure}[t]
  \centering
  \includegraphics[width=.5\linewidth]{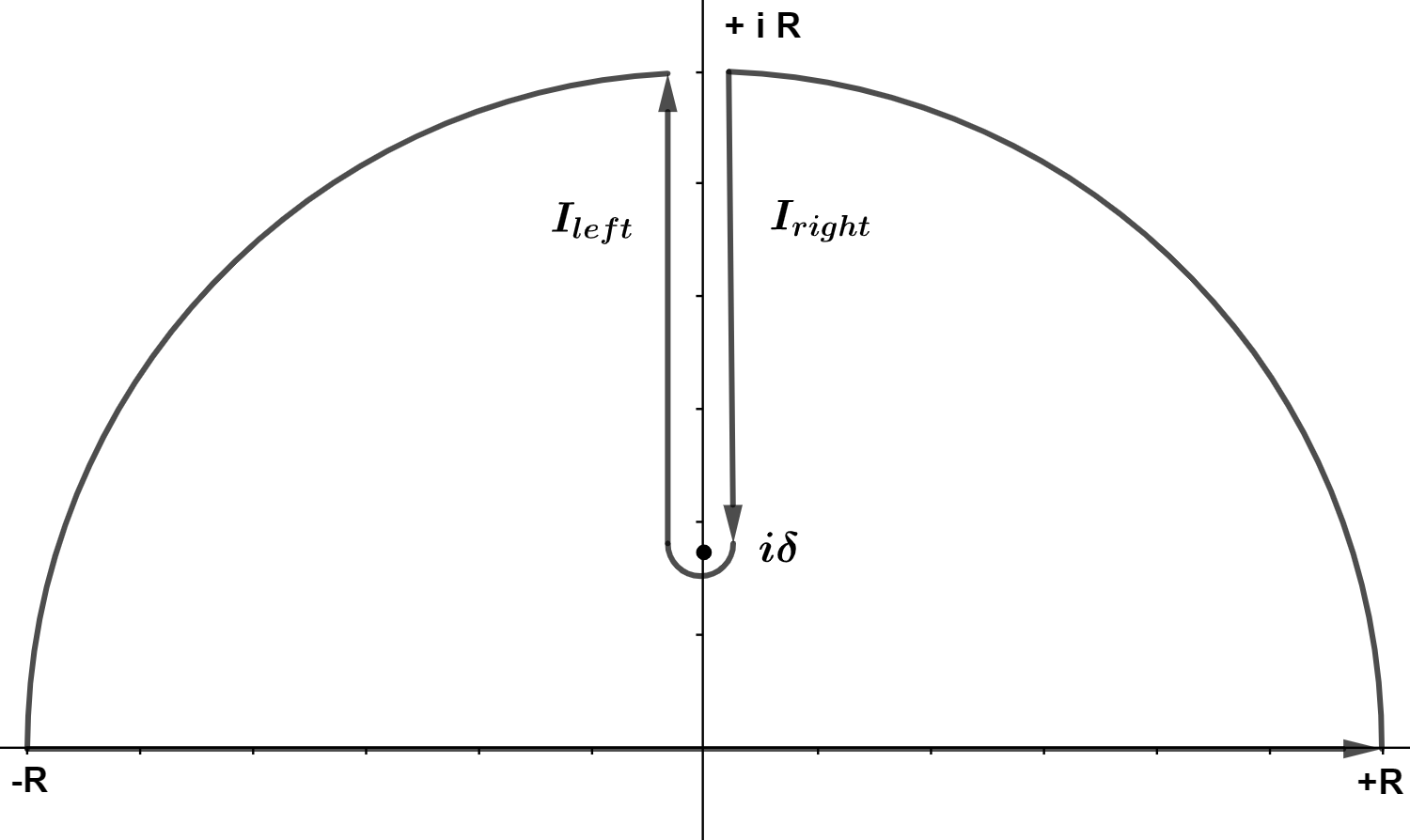}
  \caption{Key-Hole contour, with branch  cut at $i\delta$}
  \label{keyhole}
\end{figure}
The integral has a branch cut at $q_{z} = \pm i \sqrt{\frac{m^{2}}{v_{F}\beta \left(1 - \beta\right)} + q^{2}_{\perp}}$. Likewise, it has a poles at $q_{z} = \pm iq_{\perp}$.Therefore, the contour is chosen in such a way that the integral runs from $-\infty$ to $\infty$ through real line and the branch cut is from $i\sqrt{\frac{m^{2}}{v^{2}_{F}\beta \left(1 - \beta\right)} + q^{2}_{\perp}}$ to $i\infty$. This type of Contour is called the Keyhole Contour, shown in fig.\ref{keyhole}. The integral can be split into,
\begin{eqnarray}
      \oint  &=& \int^{+\infty}_{-\infty} + \int_{R} + I_{left} + I_{right} 
\end{eqnarray}
Here $I_{right}$ is the integral through the imaginary axis just along the right side of it, while $I_{left}$ is the integral just along the left side. The integral over the curve R and $\varepsilon$ will vanish in the limit $R \rightarrow \infty$. Therefore, we are left with $I_{left} + I_{right}$ and the contour integral in the LHS, where we could use residue theorem easily to solve them. The LHS is given by,
\begin{multline}
    \oint \frac{dq_{z}}{2\pi} \frac{1}{q_{\perp}^{2} + q^{2}_{z}} \text{Log}\left[\frac{m^{2}}{\beta(1 - \beta)W^{2}} + \frac{v^{2}_{F}\left(q_{\perp}^{2} + q^{2}_{z}\right)}{W^{2}} \right]\\  = \frac{1}{2q_{\perp}} \log \frac{m^{2}}{\beta(1 - \beta)W^{2}}
\end{multline}
Now, we focus on the integrals along the branch cut. First, let us call the branch cut point as $\delta = \sqrt{\frac{m^{2}}{v^{2}_{F}\beta \left(1 - \beta\right)} + q^{2}_{\perp}}$. Then $I_{left}$ and $I_{right}$ are given by,
\begin{eqnarray}
    I_{right} &=& \int^{i\delta}_{i\infty} \frac{dq_{z}}{2\pi} \frac{1}{q_{\perp}^{2} + q^{2}_{z}} \left[ \log \left(q_{z} - i\delta \right) + 2\pi i \right] \\ 
    I_{left} &=& \int^{i\infty}_{i\delta} \frac{dq_{z}}{2\pi} \frac{1}{q_{\perp}^{2} + q^{2}_{z}} \log \left(q_{z} - i\delta \right) 
\end{eqnarray}
Adding these two, we get,
\begin{eqnarray}
    I_{right} + I_{left} &=& -2\pi i\int^{i\infty}_{i\delta} \frac{dq_{z}}{2\pi} \frac{1}{q_{\perp}^{2} + q^{2}_{z}}\nonumber \\ &=&   -\frac{1}{2q_{\perp}} \log \frac{\
    \frac{\delta}{q_{\perp}} + 1}{\frac{\delta}{q_{\perp}} - 1}
\end{eqnarray}
where $\frac{\delta}{q_{\perp}} = \sqrt{\left(\frac{m}{v_{F} q_{\perp}}\right)^{2} \frac{1}{\beta \left(1 - \beta\right)} + 1}$.

This result defines the bare theory of the surface Fermions interacting via Coulomb field. Since the UV cut-off of surface theory is the bulk mass $m$, we can set the in-plane momentum $v_{F}q_{\perp} \approx m$. In this limit, the constant $\frac{\delta}{q_{\perp}} \approx \sqrt{\frac{1}{\beta \left(1 - \beta\right)} + 1}$. Thus the branch cut integral becomes,
\begin{eqnarray}
 I_{right} + I_{left} &\approx& -\frac{1}{2q_{\perp}} \log \frac{\sqrt{\frac{1}{\beta \left(1 - \beta\right)} + 1} + 1}{\sqrt{\frac{1}{\beta \left(1 - \beta\right)} + 1} - 1}
\end{eqnarray}
The first order correction to the surface Coulomb propagator as a result of the bulk electron renormalization attains the asymptotic form,
\begin{eqnarray}
    q_{\perp}\mathcal{K}^{2D(1)}_{cou}(v_{F}\textbf{q}_{\perp})|_{v_{F}\textbf{q}_{\perp}\approx m} &\approx& \frac{e^{2}_{b0}}{2}\frac{e^{2}_{b0}}{6\pi^{2}v_{F}} \left[\log \frac{m}{ W} + 1.2611 \right]\nonumber\\
\end{eqnarray}
where the number 1.2611 is a result of the following integration,
\begin{eqnarray}
1.2611 &=& 3\int^{1}_{0} d\beta \beta(1 - \beta)\biggl[ \log \frac{\sqrt{\frac{1}{\beta \left(1 - \beta\right)} + 1} + 1}{\sqrt{\frac{1}{\beta \left(1 - \beta\right)} + 1} - 1}\nonumber\\ &-& \log \beta(1 - \beta) \biggr] 
\end{eqnarray}
\subsection*{Evaluation of $\mathcal{K}^{2D(1)}_{e-ph}(\textbf{q}_{\perp})$}

It is to be noted that unlike the Coulomb integral, we must include the regularization term  since the integral is UV divergent. This is evident from the expression for effective phonon propagator in the 3D bulk given in
Eq.\ref{phononprop3D}, whose dimension is zero. Therefore, the first order quantum correction to the surface phonon propagator has the form,
\begin{widetext}
\begin{eqnarray}
\mathcal{K}^{2D(1)}_{e-ph}(\textbf{q}_{\perp}) &=& \frac{e^{2}_{b}(\Omega_{D})}{\pi^{2}v_{F}}\int \frac{dq_{z}}{2\pi} \left(\frac{M^{2}}{q_{z}^{2} + M^{2}}\right)^{2} \frac{v^{2}_{p} \left( q_{\perp}^{2} + q^{2}_{z}\right)}{\omega^{2} - v^{2}_{p}\left(q_{\perp}^{2} + q^{2}_{z}\right) } \Pi(v_{F}q_{\perp}, v_{F}q_{z}) \\ &=& \frac{e^{2}_{b0}(\Omega_{D})}{\pi^{2}v_{F}}\int \frac{dq_{z}}{2\pi} \left(\frac{M^{2}}{q_{z}^{2} + M^{2}}\right)^{2} \frac{v^{2}_{p} \left( q_{\perp}^{2} + q^{2}_{z}\right)}{\omega^{2} - v^{2}_{p}\left(q_{\perp}^{2} + q^{2}_{z}\right) }\int^{1}_{0} d\beta \, \beta \left(1 - \beta\right) \text{Log}\left[\frac{m^{2}}{\beta(1 - \beta)\Omega_{D}^{2}} + \frac{v^{2}_{F}\left(q_{\perp}^{2} + q^{2}_{z}\right)}{\Omega_{D}^{2}}\right]\nonumber \\ &=& \frac{e^{2}_{b0}(\Omega_{D})}{\pi^{2}v_{F}}\int \frac{dq_{z}}{2\pi} \left(\frac{M^{2}}{q_{z}^{2} + M^{2}}\right)^{2} \frac{v^{2}_{p} \left( q_{\perp}^{2} + q^{2}_{z}\right)}{\omega^{2} - v^{2}_{p}\left(q_{\perp}^{2} + q^{2}_{z}\right) }\int^{1}_{0} d\beta \, \beta \left(1 - \beta\right) \text{Log} \frac{m^{2}}{\beta(1 - \beta)\Omega_{D}^{2}}\nonumber \\ &+& \frac{e^{2}_{b0}(\Omega_{D})}{\pi^{2}v_{F}}\int \frac{dq_{z}}{2\pi} \left(\frac{M^{2}}{q_{z}^{2} + M^{2}}\right)^{2} \frac{v^{2}_{p} \left( q_{\perp}^{2} + q^{2}_{z}\right)}{\omega^{2} - v^{2}_{p}\left(q_{\perp}^{2} + q^{2}_{z}\right) }\int^{1}_{0} d\beta \, \beta \left(1 - \beta\right) \text{Log} \left[1 + \frac{v^{2}_{F}\beta(1 - \beta) \left(q^{2}_{\perp} + q^{2}_{z} \right)}{m^{2}} \right]    
\end{eqnarray}
\end{widetext}

In the last step, we have written the $\log$ expression as a sum of two terms for reasons that will be clear soon.
Both the terms have poles in the Upper Half Plane at $q_{z} =  i\sqrt{v^{2}_{p}q_{\perp}^{2} - \omega^{2}}, \pm iM$. While the first term is smooth except for the poles in the Complex plane, the second term has a branch cut at $q_{z} =  i \sqrt{\frac{m^{2}}{v^{2}_{F}\beta \left(1 - \beta\right)} + q^{2}_{\perp}}$.
 Solving the first term will give a logarithmic correction to the bare propagator, with the log term being a ratio of the UV cut off scales for the surface and the bulk, given by $m/\Omega_{D}$. The second term could be solved only by using a branch cut integral just as in the previous case. Let us call the integral expression to be $I$,
\\
\\
\begin{multline}
      I = \int \frac{dq_{z}}{2\pi} \left(\frac{M^{2}}{q_{z}^{2} + M^{2}}\right)^{2} \frac{v^{2}_{p} \left( q_{\perp}^{2} + q^{2}_{z}\right)}{\omega^{2} - v^{2}_{p}\left(q_{\perp}^{2} + q^{2}_{z}\right) } \\ \log \left[1 + \frac{v^{2}_{F}\beta(1 - \beta) \left(q^{2}_{\perp} + q^{2}_{z} \right)}{m^{2}} \right] \label{I}
\end{multline}

Consider the same Keyhole Contour given in fig.\ref{keyhole}. The integral over the curve R will vanish in the limit $R \rightarrow \infty$ The LHS is just the sum of the residues of the integrand at $q_{z} = i\sqrt{q^{2}_{\perp} - \frac{\omega^{2}}{v^{2}_{p}}}$ and $q_{z} = iM\sqrt{1 - \epsilon}$ which is given by,
\begin{widetext}
\begin{eqnarray}
\mathcal{C} &=& \oint \frac{dq_{z}}{2\pi} \left(\frac{M^{2}(1 - \epsilon)}{q_{z}^{2} + M^{2}(1 - \epsilon)}\right)^{2} \frac{v^{2}_{p} \left( q_{\perp}^{2} + q^{2}_{z}\right)}{\omega^{2} - v^{2}_{p}\left(q_{\perp}^{2} + q^{2}_{z}\right) }  \log \left[1 + \frac{v^{2}_{F}\beta(1 - \beta) \left(q^{2}_{\perp} + q^{2}_{z} \right)}{m^{2}} \right] \\ &=& \frac{M}{8} \left(\frac{2q^{2}_{\perp}}{\frac{\omega^{2}}{v^{2}_{p}} - q^{2}_{\perp}} - \frac{\frac{\omega^{2}}{v^{2}_{p}}}{\frac{\omega^{2}}{v^{2}_{p}} - q_{\perp}^{2}}\left( \frac{M^{2}}{\left(M - \sqrt{q_{\perp}^{2} - \frac{\omega^{2}}{v^{2}_{p}}}\right)^{2}} + \frac{M^{2}}{\left(M + \sqrt{q_{\perp}^{2} - \frac{\omega^{2}}{v^{2}_{p}}}\right)^{2}}\right) \right)\log \left[1 + \frac{v^{2}_{F}\beta(1 - \beta)}{m^{2}}\left( q_{\perp}^{2} - M^{2}\right)\right]\nonumber\\ &+& \frac{M^{3}}{8} \frac{\frac{\omega^{2}}{v^{2}_{p}}}{\frac{\omega^{2}}{v^{2}_{p}} - q_{\perp}^{2}}\left(\frac{1}{\left(M - \sqrt{q_{\perp}^{2} - \frac{\omega^{2}}{v^{2}_{p}}}\right)^{2}} - \frac{1}{\left(M + \sqrt{q_{\perp}^{2} - \frac{\omega^{2}}{v^{2}_{p}}}\right)^{2}} \right)\log \left[1 + \frac{v^{2}_{F}\beta(1 - \beta)}{m^{2}}\frac{\omega^{2}}{v^{2}_{p}} \right]\nonumber \\  &+& \frac{M}{4}\left(\frac{2q^{2}_{\perp}}{\frac{\omega^{2}}{v^{2}_{p}} - q^{2}_{\perp}} - \frac{\frac{\omega^{2}}{v^{2}_{p}}}{\frac{\omega^{2}}{v^{2}_{p}} - q_{\perp}^{2}}\left(\frac{M}{M + \sqrt{q_{\perp}^{2} - \frac{\omega^{2}}{v^{2}_{p}}}} + \frac{M}{M - \sqrt{q_{\perp}^{2} - \frac{\omega^{2}}{v^{2}_{p}}}}\right) \right)\frac{4\beta(1 - \beta)}{1 + \frac{v^{2}_{F}\beta(1 - \beta)}{m^{2}}\left( q_{\perp}^{2} - M^{2}\right)} \label{residue_ph}
\end{eqnarray}
\end{widetext}
And the sum of the integrals along the branch cut, $I_{left} + I_{right}$ is given by,
\begin{widetext}
\begin{eqnarray}
I_{left} + I_{right} &=&
-2\pi i\int^{i\infty}_{i\delta} \frac{dq_{z}}{2\pi} \left(\frac{M^{2}}{q_{z}^{2} + M^{2}}\right)^{2} \frac{ q_{\perp}^{2} + q^{2}_{z}}{\frac{\omega^{2}}{v^{2}_{p}} - \left(q_{\perp}^{2} + q^{2}_{z}\right)} \\ &=&
\frac{M^{4}}{\frac{\omega^{2}}{v^{2}_{p}} - q^{2}_{\perp} + M^{2}}\biggl[ -\frac{\omega^{2}}{\frac{\omega^{2}}{v^{2}_{p}} - q^{2}_{\perp} + M^{2}}\left(\frac{1}{2M} \log \frac{\delta - M}{\delta + M} + \frac{1}{2\sqrt{q^{2}_{\perp} -\frac{\omega^{2}}{v^{2}_{p}}}} \log \frac{\delta - \sqrt{q^{2}_{\perp} -\frac{\omega^{2}}{v^{2}_{p}}}}{\delta + \sqrt{q^{2}_{\perp} -\frac{\omega^{2}}{v^{2}_{p}}}} \right)\nonumber\\ &+& \frac{M^{2} - q^{2}_{\perp}}{2M}\left(\frac{1}{2M^{2}}\log \frac{\delta + M}{\delta - M} - \frac{\delta}{M}\frac{1}{\delta^{2} - M^{2}} \right)\biggr]  
\end{eqnarray}
\end{widetext}
The resulting solution will define the first order quantum correction to the bare theory of the surface Fermions interacting via lattice phonon field. Since the UV cut-off of surface theory is the bulk mass $m$, we can set the in-plane momentum $v_{F}q_{\perp} \approx m$. Also the phonon frequency is set to $\omega \approx 0$, the limit which supports Superconducting pairing. In this limit, the sum of the residues of the keyhole contour, whose full expression is given in Eq.\ref{residue_ph} attains the limiting value,
\begin{eqnarray}
\mathcal{C} &\approx& -\frac{M}{4} \log \left[1 - 3\beta(1 - \beta) \right] - \frac{M}{4} \frac{8\beta(1 - \beta)}{1 - 3\beta(1 - \beta)}\nonumber\\
\end{eqnarray}
Similarly, the sum of the integrals along the two sides of the branch cut become,
\begin{multline}
I_{right} + I_{left} = \\ \frac{M}{4} \left[\log \frac{\frac{1}{2}\sqrt{\frac{1}{\beta(1 - \beta)} + 1} + 1}{\frac{1}{2}\sqrt{\frac{1}{\beta(1 - \beta)} + 1} - 1} - \frac{\sqrt{\frac{1}{\beta(1 - \beta)} + 1}}{\frac{1}{4\beta(1 - \beta)} + \frac{1}{4} - 1} \right]    
\end{multline}
The solution of the integral denoted by $I$ given in eqn \ref{I} is obtained by subtracting the integral along the branch cuts from the sum of the residues. Thus, we have,
\begin{eqnarray}
I &=& \mathcal{C} - I_{left} - I_{right}\nonumber \\ &=&  -\frac{M}{4} \biggl[ 2\log\left[2 + \sqrt{1 + \frac{1}{\beta(1 - \beta)}} \right] + \log \beta(1 - \beta)\nonumber \\ &-&\frac{2}{1 + \frac{1}{2}\sqrt{\frac{1}{\beta(1 - \beta)} + 1}} \biggr]
\end{eqnarray}
Finally, let us write down the first order correction to the surface phonon propagator as a result of the bulk electron renormalization,
\begin{multline}
\mathcal{K}_{e-ph}^{2D(1)}(\omega, v_{F}q_{\perp})|_{\omega \approx 0, v_{F}q_{\perp} = m}\approx \\ - \frac{G^{2}_{fp0}m}{2v_{F}} \frac{e^{2}_{b}(\Omega_{D})}{3\pi^{2}v_{F}} \left[\log \frac{m}{\Omega_{D}} + 1.06874.. \right]
\end{multline}
where the numerical constant 1.06874 is the result of the following integration,
\begin{eqnarray}
1.06874 &=& 6 \int^{1}_{0} d\beta \beta(1 - \beta) \biggl[\log\left[2 + \sqrt{1 + \frac{1}{\beta(1 - \beta)}} \right]\nonumber\\ &-& \log \sqrt{\beta(1 - \beta)}\nonumber  -\frac{2}{1 + \frac{1}{2}\sqrt{\frac{1}{\beta(1 - \beta)} + 1}} \biggr]
\end{eqnarray}


\begin{thebibliography}{0}%
\makeatletter
\providecommand \@ifxundefined [1]{%
 \@ifx{#1\undefined}
}%
\providecommand \@ifnum [1]{%
 \ifnum #1\expandafter \@firstoftwo
 \else \expandafter \@secondoftwo
 \fi
}%
\providecommand \@ifx [1]{%
 \ifx #1\expandafter \@firstoftwo
 \else \expandafter \@secondoftwo
 \fi
}%
\providecommand \natexlab [1]{#1}%
\providecommand \enquote  [1]{``#1''}%
\providecommand \bibnamefont  [1]{#1}%
\providecommand \bibfnamefont [1]{#1}%
\providecommand \citenamefont [1]{#1}%
\providecommand \href@noop [0]{\@secondoftwo}%
\providecommand \href [0]{\begingroup \@sanitize@url \@href}%
\providecommand \@href[1]{\@@startlink{#1}\@@href}%
\providecommand \@@href[1]{\endgroup#1\@@endlink}%
\providecommand \@sanitize@url [0]{\catcode `\\12\catcode `\$12\catcode
  `\&12\catcode `\#12\catcode `\^12\catcode `\_12\catcode `\%12\relax}%
\providecommand \@@startlink[1]{}%
\providecommand \@@endlink[0]{}%
\providecommand \url  [0]{\begingroup\@sanitize@url \@url }%
\providecommand \@url [1]{\endgroup\@href {#1}{\urlprefix }}%
\providecommand \urlprefix  [0]{URL }%
\providecommand \Eprint [0]{\href }%
\providecommand \doibase [0]{http://dx.doi.org/}%
\providecommand \selectlanguage [0]{\@gobble}%
\providecommand \bibinfo  [0]{\@secondoftwo}%
\providecommand \bibfield  [0]{\@secondoftwo}%
\providecommand \translation [1]{[#1]}%
\providecommand \BibitemOpen [0]{}%
\providecommand \bibitemStop [0]{}%
\providecommand \bibitemNoStop [0]{.\EOS\space}%
\providecommand \EOS [0]{\spacefactor3000\relax}%
\providecommand \BibitemShut  [1]{\csname bibitem#1\endcsname}%
\let\auto@bib@innerbib\@empty
\end{thebibliography}%


\begin{thebibliography}{References}



\bibitem{Wen1}X. G. Wen Phys. Rev. B 41, 12838 (1990).

\bibitem{Wen2}X. G. Wen, Advances in Physics, 44, 405 (1995).

\bibitem{Chang}A. M. Chang,   M. K. Wu, C. C. Chi, L. N. Pfeiffer, K. W. West, Phys. Rev. Lett. 86, 143--146, (2001).



\bibitem{Kane05} C. L. Kane and E. J. Mele, Phys. Rev. Lett. {\bf95} 146802 (2005); Phys. Rev. Lett. {\bf 95}, 226801 (2005).

\bibitem{Bernevig06a} B. A. Bernevig and S. C. Zhang, Phys. Rev. Lett. 96, 106802 (2006).

\bibitem{Bernevig06b} B. A. Bernevig, T. L. Hughes, and S. C. Zhang, Science {\bf314}, 1757 (2006).

\bibitem{Fu07} L. Fu, C. L. Kane and E. J. Mele, Phys. Rev. Lett. {\bf98}, 106803 (2007). 

\bibitem{Moore07} J. E. Moore and L. Balents, Phys. Rev. B {\bf75} 121306(R) (2007).

\bibitem{Qi10a} X. L. Qi, T. L. Hughes, and S. C. Zhang, Phys. Rev B {\bf82}, 184516 (2010).

\bibitem{Hasan10} M. Z. Hasan and C. L. Kane, Rev. Mod. Phys. {\bf82}, 3045 (2010).

\bibitem{Qi11} X.-L. Qi and S.-C. Zhang, Rev. Mod. Phys. {\bf83}, 1057 (2011).

\bibitem{Bernevig} B. A. Bernevig and T. L. Hughes, {\it Topological insulators and topological superconductors} (Princeton University Press, 2013).


\bibitem{Fu08} L. Fu and C. L. Kane, Phys. Rev. Lett. {\bf100}, 096407 (2008).

\bibitem{Lutchyn10}R. M. Lutchyn, J. D. Sau, and S. D Sarma, Phys. Rev. Lett. {\bf105}, 077001 (2010).

\bibitem{Lutchyn11} J. D. Sau,  R. M. Lutchyn, S. Tewari,  S. D. Sarma, Phys. Rev. Lett. 040502 (2010).

\bibitem{Fidkowski}L. Fidkowski, X. Chen and A. Vishwanath, Phys. Rev.X 3, 041016 (2013)

\bibitem{Metlitski} M. Metlitski, C. L. Kane and M. P. A. Fisher, Phys. Rev.B 92, 12511 (2015).

\bibitem{Wang1} C. Wang and T. Senthil, Phys. Rev. X 5, 041031 (2015).

\bibitem{Wang2} C. Wang, A. Potter and T. Senthil, Science 343, 6171 (2014).

\bibitem{Song}H. Song, S. J. Huang, L. Fu and M. Hermele, Phys. Rev.X 7, 011020 (2017).

\bibitem{Neupert}T. Neupert, S. Rachel, R. Thomale, and M. Greiter (2015), Phys. Rev. Lett. 115, 017001.

\bibitem{Wu}C. Wu , B. A. Bernevig, S. C. Zhang, Phys. Rev. Lett. 96, 106401 (2006).


\bibitem{Peshkin} M. E. Peshkin and D. V. Schroeder, {\it An Introduction to Quantum Field Theory} (Addison-Willey Publishing Company, 1995).

\bibitem{gonzalez99}J. Gonz\'alez, F. Guinea, M. A. H. Vozmediano, Phys. Rev. B. 59, R2474--R2477 (1999).

\bibitem{schmalian07}D. E. Sheehy, J. Schmalian, Phys. Rev. Lett. 99, 226803 (2007).

\bibitem{sdsharma}S. D. Sarma,  E. H. Hwang, W. K. Tse, Phys. Rev. B 75, 121406 (2007).

\bibitem{lucas}A. Lucas, K. C. Fong, J. Phys. Condens. Matter, 053001 (2018).

\bibitem{abrikosov}A. A. Abrikosov, A. Beknazarov, {\it Fundamentals of the Theory of Metals}(North-Holland, 1988). 
\bibitem{Ludwig}A. P. Schnyder, S. Ryu, A. Furusaki, and A. W. W. Ludwig, Phys. Rev. B 78, 195125 (2008).

\bibitem{QiZhang} X. L. Qi,  T. L. Hughes,  S. Raghu, S. C. Zhang, Phys. Rev. Lett. 102, 187001, (2009); X. L. Qi, T. L. Hughes, S. C. Zhang, Phys. Rev. B
81, 134508 (2010).

\bibitem{kitaev}A Kitaev, AIP Conference Proceedings 1134, 22-30 (2009).

\bibitem{Lee07} S. S. Lee, Phys. Rev. B {\bf76}, 075103 (2007).

\bibitem{Grover14} T. Grover, D. N. Sheng and A. Vishwanath, Science {\bf344}, 280 (2014).

\bibitem{Ponte14} P. Ponte and S. S. Lee, New J. Phys. 16, 013044 (2014).


\bibitem{Li17} Z. X. Li, Y. F. Jiang, and H. Yao, Phys. Rev. Lett. 119, 107202 (2017).


\bibitem{Li18} Z. X. Li, A. Vaezi, C. B. Mendl and H Yao, Science Advances. 4, 11 (2018).

\bibitem{Jian17} S. K. Jian, C. H. Lin, J. Maciejko, and H. Yao, Phys. Rev. Lett. {\bf 118}, 166802 (2017).
















\bibitem{Jian15} S.K. Jian, Y. F. Jiang and H. Yao, Phys. Rev. Lett. 114, 237001 (2015).

\bibitem{Sachdev} S. Sachdev, Quantum Phase Transitions (Cambridge University Press, 1999).



\end{thebibliography}
\end{document}